\documentclass[10pt,journal,twocolumn]{IEEEtran}

\usepackage{subcaption,wrapfig,graphicx,booktabs,fancyhdr,amsmath,amsfonts,tikz}
\usepackage{cite,bm,amssymb,amsthm,url,multirow,times,enumitem,comment,multicol,textcomp}
\usepackage[linesnumbered,ruled]{algorithm2e}

\DeclareMathAlphabet{\mathpzc}{OT1}{pzc}{m}{it}
\usepackage[detect-weight,binary-units]{siunitx}
\usepackage{hhline}
\usepackage[usestackEOL]{stackengine}
\usepackage{letltxmacro}
\usepackage{bbm}
\usepackage[colorlinks=true, linkcolor=black, citecolor=blue,urlcolor=blue]{hyperref}

\DeclareSIUnit[number-unit-product = ]\percent{\char`\%}
\DeclareSIUnit\people{people}

\ifCLASSOPTIONdraftcls
\makeatletter
\LetLtxMacro{\oldincludegraphics}{\includegraphics}
\DeclareRobustCommand\includegraphics{\@ifnextchar[\@@includegraphics\oldincludegraphics}
\newcommand\@@includegraphics[2][]{\oldincludegraphics{#2}}
\makeatother
\fi

\newcommand{\errorrange}{\SI{19}{cm}\\\SI{25}{cm}}

\newcommand{\orbitclockrms}{\SI{14}{cm}}

\makeatletter
\newcommand{\dummylabel}[2]{\def\@currentlabel{#2}\label{#1}}
\makeatother

\newcommand{\pointingThroughputPenaltyA}{\SI{0.10}{\percent}}
\newcommand{\pointingTsteermaxA}{\SI{260}{\milli\second}}
\newcommand{\pointingThroughputPenaltyB}{\SI{0.5}{ppm}}

\newcommand{\n}{5}
\newcommand{\Ncells}{850000}
\newcommand{\Nsats}{10000}
\newcommand{\Nchannels}{76}
\newcommand{\Nbeams}{15}
\newcommand{\Nbc}{264}
\newcommand{\Nadj}{6}
\newcommand{\W}{\SI{50}{\MHz}}
\newcommand{\D}{\SI{29}{\kilo\meter}}
\newcommand{\Tburst}{\SI{500}{\micro\second}}
\newcommand{\Tsetup}{\SI{5}{\milli\second}}
\newcommand{\Tswitch}{\SI{100}{\micro\second}}
\newcommand{\Tperiod}{\SI{1}{\second}}

\newcommand{\dpnt}{\SI{0.33}{\percent}}
\newcommand{\dUTULMax}{\SI{99.67}{\percent}}
\newcommand{\minElevation}{\SI{40}{\degree}}
\newcommand{\Tquantize}{\SI{1}{\micro\second}}
\newcommand{\TXreservationMax}{\SI{1.60}{\percent}}
\newcommand{\RXreservationMax}{\SI{0.03}{\percent}}
\newcommand{\RXremaining}{\SI{99.97}{\percent}}
\newcommand{\downlinkReservationMax}{\SI{1.60}{\percent}}
\newcommand{\downlinkReservationRound}{\SI{1.6}{\percent}}
\newcommand{\downlinkRemaining}{\SI{98.4}{\percent}}
\newcommand{\setupReservationMax}{\SI{11.3}{\percent}}

\newcommand{\powerReservationMax}{\SI{0.77}{\percent}}
\newcommand{\serviceLatMax}{\SI{60}{\degree}}
\newcommand{\PARmin}{\SI{9.6}{}}
\newcommand{\broadbandEquivalentService}{\SI{5.7}{Mbps}}

\newcommand{\broadbandSteeringRate}{\SI{0.73}{\degree\per\second}}

\newcommand{\assignmentUplinkCost}{\SI{54}{\mebi\byte}}
\newcommand{\uplinkRatePerSV}{\SI{3.5}{kbps}}
\newcommand{\assignmentSizeBits}{59}
\newcommand{\PFD}{\SI{-122.0}{\dB(\W\per\meter^2\per\MHz)}}

\newcommand{\cellAreaMax}{\SI{1090}{\km^2}}

\newcommand{\SoPCoverageRatio}{6.3}
\newcommand{\expectedComplexityMax}{\SI{4.4e+06}{\;steps}}
\newcommand{\Rturbo}{\SI{114.5}{Mbps}}
\newcommand{\beamFWHM}{\SI{2.0}{\degree}}

\newcommand{\pmax}{\SI{92.7}{\people\per\km^2}}
\newcommand{\pmaxDeveloped}{\SI{63.2}{\people\per\km^2}}

\newcommand{\serviceRegionPopFraction}{\SI{99.8}{\percent}}

\usepackage{afterpage}
\usepackage{balance}

\fancypagestyle{versionfooter}{
    \fancyfoot[L]{v2.0
}
    \fancyfoot[C]{\thepage}
}

\newcommand\Ku{\ensuremath{\mathrm{K}_\mathrm{u}}}
\newcommand\Ka{\ensuremath{\mathrm{K}_\mathrm{a}}}

\begin{document}
\title{Fused Low-Earth-Orbit GNSS}

\author{
\IEEEauthorblockN{Peter~A.~Iannucci and Todd~E.~Humphreys} \\
\IEEEauthorblockA{\textit{Radionavigation Laboratory} \\
\textit{The University of Texas at Austin}\\
Austin, TX, USA}\\
peter.iannucci@austin.utexas.edu, todd.humphreys@utexas.edu
}

\maketitle
\begin{abstract}Traditional Global Navigation Satellite System (GNSS) immunity to interference
may be approaching a practical performance ceiling.  Greater gains are possible
outside traditional GNSS orbits and spectrum. GNSS from low Earth orbit (LEO)
has long been viewed as promising but expensive, requiring large constellations
for rapid navigation solutions.  The recent emergence of commercial broadband
LEO mega-constellations invites study on dual-purposing these for both
communications---their primary mission---and a secondary positioning,
navigation, and timing (PNT) service.  Operating at shorter wavelengths than
traditional GNSS, these constellations would permit highly directive,
relatively compact receiver antennas.  PNT-specific on-orbit resources would
not be required: the transmitters, antennas, clocks, and spectrum of the
hosting broadband network would suffice for PNT.  Non-cooperative use of LEO
signals for PNT is an option, but cooperation with the constellation operator
(``fusion'' with its communications mission) eases the burden of tracking a
dense, low-altitude constellation from the ground and enables a receiver to
produce single-epoch stand-alone PNT solutions. This paper proposes such a
cooperative concept, termed fused LEO GNSS.  Viability hinges on opportunity
cost, or the burden a secondary PNT mission imposes on the communications
constellation operator. This is assessed in terms of time-space-bandwidth
product and energy budget.  It is shown that a near-instantaneous-fix PNT
service over $\pm\serviceLatMax$ latitude (covering \serviceRegionPopFraction\
of the world's population\nocite{GPWv4}) with positioning performance superior
to traditional GNSS pseudoranging would cost less than
\downlinkReservationRound\ of downlink capacity for the largest of the new
constellations, SpaceX's Starlink.  This allocation is comparable to adding one
user consuming \broadbandEquivalentService\ of broadband service to each cell.

\end{abstract}


\pagestyle{plain}
\thispagestyle{versionfooter}

\section{Introduction}
Use of low-Earth orbit (LEO) constellations for positioning, navigation, and
timing (PNT) dates back to the earliest operational satellite navigation
constellation, TRANSIT \cite{morton2020position}.  Based on Doppler
measurements extracted from narrowband UHF signals received from a single
satellite at a time, TRANSIT required several minutes for convergence to a
sub-100-meter solution.

The trade studies from which the Global Positioning System (GPS) was later
conceived revealed that a medium Earth orbit (MEO) system with wideband signals
would be more resistant to jamming than TRANSIT and would be capable of
satellite-redundant instantaneous positioning with only a few dozen space
vehicles (SVs) \cite[Ch. 1]{morton2020position}.  L band was chosen because its
wavelengths are short enough for ionospheric transparency, yet long enough to
avoid significant attenuation due to rainfall and water vapor
\cite{j_spilker96_atp,j_spilker96_int,teunissen2017springer}.  By now all
traditional global navigation satellite systems (GNSS) have settled into a
system architecture similar to that of GPS, to great success: billions of users
across the globe benefit from low-cost, high-accuracy, near-instantaneous
positioning and timing.

Nevertheless, the traditional GNSS architecture suffers from some
deficiencies. Non-GNSS uses of the congested space-to-Earth spectrum in L band
have prevented allocation of much greater bandwidth for GNSS in that band.
Constellation survivability is limited by the small number of SVs, which make
attractive targets for anti-satellite warfare \cite{dolman2012new,
  washTimesRaymond2021}.  Jamming immunity is limited by the weakness of the
signals, which, being diffused over an entire hemisphere, are easily
overwhelmed \cite{humphreysGNSShandbook,psiakiNewBlueBookspoofing}.  And
positioning precision is limited by both signal weakness and bandwidth,
which place information-theoretic lower bounds on ranging uncertainties
\cite{j_spilker96_atp}.

In response to a pressing need for greater robustness and accuracy, GNSS has
evolved over the past two decades.  Several new constellations have been
launched, and new signals have been introduced at separate frequencies---most
with binary offset carrier waveforms that more efficiently allocate signal
power \cite{teunissen2017springer, morton2020position}.  Nonetheless, GNSS
remains principally MEO, L-band, and confined to a bandwidth occupying less
than \SI{125}{MHz}.  Given tight budgets and enormous design inertia owing to
the need for backward compatibility, radical changes in traditional GNSS over
the next 30 years are unlikely.  Spot beams, a promising feature of the GPS III
program for improved jamming immunity \cite{luba2005gps}, have been abandoned.
Calls to introduce new GNSS signals in C band (e.g., \cite{t_humphreys08_cfi})
have not gained traction.  Upgraded SVs and more sophisticated receiver
antennas will continue to extract gains in interference immunity, but likely
not tens of decibels.

In short, traditional GNSS have been brilliantly successful, yet for some
applications they remain inadequate with regard to accuracy, constellation
survivability, or robustness to interference---for both civil and military
users.  To address these limitations, this paper introduces a concept for LEO
PNT that exploits current and upcoming broadband LEO mega-constellations via a
novel ``fused'' communications-and-PNT service.  The practical costs and
challenges facing past LEO GNSS proposals, including hosted-payload LEO GNSS
and signal-of-opportunity (SoP) LEO GNSS, motivate the paper's proposed
architecture.

This paper makes three primary contributions.  First, it summarizes the
features of modern broadband LEO system design and operation relevant to
dual-purposing such systems for PNT.  Second, it presents a detailed concept of
operations for {\em fused LEO GNSS}, to be defined in the next section.  Third,
it provides an analysis of the opportunity cost to constellation providers for
re-allocating resources to provide a fused PNT service. The paper is organized
around these primary contributions.

The earlier paper published in \cite{iannucci2020fused} is complementary to the
present paper, which provides a complete description of the fused LEO GNSS
concept and a detailed opportunity cost analysis.  The reader is referred to
\cite{iannucci2020fused} for analyses of achievable fused LEO GNSS positioning
precision and anti-jam advantage compared to traditional GNSS.  Summary values
from these analyses are provided in Table \ref{tab:hostedvsfused} for reference.

\section{LEO GNSS}
Expansion of GNSS back to the LEO ambit of TRANSIT beckons as a promising way
to address the limitations of traditional GNSS.  Mega-constellations of
commercial satellites in LEO are being launched (SpaceX's Starlink and OneWeb's
constellations) or planned (Amazon's Kuiper constellation) to provide broadband
connectivity across the globe.  Such services' global reach, low latency, and
wide bandwidth situate them to revolutionize broadband communications.

This paper seeks to establish a less-obvious assertion: These constellations
could also revolutionize satellite-based PNT.  Their SVs are far nearer and
more numerous than those of traditional GNSS in MEO or geostationary orbit, and
their communications transponders have both exceedingly high gain and access to
a vast allocation of spectrum.  Potential commercial LEO PNT signals are thus
more precise, powerful, and jam-resistant than those of traditional GNSS.

Dual-purposing LEO communications constellations for PNT is not a new concept.
The emergence of the Globalstar and Iridium constellations in the late 1990s
offered the prospect of LEO-provided navigation based on both Doppler and
ranging.  These constellations employ communications waveforms whose frequency
and group delay can be measured opportunistically (i.e., without special
cooperation by the constellation operator) and converted to typical GNSS
observables: Doppler, phase, and pseudorange measurements
\cite{levanon1998quick,rabinowitz2000some,lawrence2016test,khalife2019receiver,
  khalife2020blind,kassas2020navigation,kassas2020leo} (see
\cite[Ch. 2]{morton2020position} for definitions of these observables).  But as
with TRANSIT, only one or two Globalstar or Iridium SVs are simultaneously
visible to a typical terrestrial user, preventing accurate instantaneous
positioning.  Instead, both theoretical \cite{levanon1984theoretical} and
experimental
\cite{lawrence2016test,lawrence2017navigation,ardito2019performance,kassas2020leo}
research has shown that several minutes of single-satellite passage across the
sky are necessary for positioning to an accuracy below 100 meters.  This remains
true for IridiumNEXT, whose constellation is patterned after the original
Iridium constellation \cite{benzerrouk2019alternative}.

The emergence of mega-constellations of LEO satellites whose signals can be
exploited for many-in-view navigation, whether opportunistically or with the
cooperation of the constellation operator, is an entirely new phenomenon.  The
literature exploring use of such constellations for PNT begins with
\cite{Reid2016LeveragingCB,Reid2018BroadbandLEO}.  The current paper belongs in
this category.

Although not originally intended for PNT, broadband mega-constellations are
designed for rapid technological refresh via software or hardware, and so may
be adaptable for PNT.  But unlike traditional GNSS, in which costs are borne by
nation-states and service is free-of-charge, commercial GNSS providers will
seek to recoup costs from users.  For such a scheme to be viable, it must be
{\em economical}: that is, it must offer fundamental advantages over
traditional (free) GNSS commensurate with the price tag, otherwise there will
be no demand; and must be sufficiently inexpensive to provide, otherwise there
will be no supply.  This paper explores both facets of this problem.

\subsection{Hosted Payload LEO GNSS}
In their groundbreaking work, Reid et al.%
\cite{Reid2016LeveragingCB,Reid2018BroadbandLEO,reid2020leo,SatNavXona2020}
analyzed the performance of potential LEO GNSS implemented using {\em hosted
  payloads}: dedicated PNT hardware onboard each satellite.  There are good
reasons to explore a hosted payload solution: Such payloads are modular,
independent of the satellite's primary communications mission, and may be
iterated and upgraded for future launches.  As laid out by Reid et al., hosted
PNT signals provide continuous global coverage and may be incorporated into
user pseudorange navigation equipment nearly as readily as traditional GNSS
signals.  Reid et al. estimate that the system would enjoy a \SI{30}{dB}
improvement in signal-to-noise ratio, and thus resistance to jamming, over
traditional GNSS.

A hosted payload approach along those lines is not radically dissimilar to
traditional GNSS.  No theoretical obstacle bars the way.  However, space
hardware development is costly and challenging as a practical matter.  And a
hosted payload would be costly: besides the cost of each payload, there are
costs associated with renting space and hookups on the host satellite, costs
for running necessary radiofrequency interference and compatibility testing,
and both costs and risks of delay in securing the necessary frequency
allocations.

\subsection{Signal-of-Opportunity LEO GNSS}
A growing area of PNT practice draws measurements from so-called signals of
opportunity (SoPs), typically wireless communications
signals\cite{kassas2014receding,kassas2014greedy,kassas2020sops}.  SoP
techniques seek to eliminate the need for cooperation with the wireless system
operator.  Satellite downlink signals from the new LEO mega-constellations
could be processed as SoPs, as has been done previously with the smaller
Iridium and Globalstar constellations
\cite{lawrence2016test,khalife2019receiver,khalife2020blind,kassas2020navigation,
  lawrence2017navigation,ardito2019performance,benzerrouk2019alternative,kassas2020leo}.
Such SoP-based LEO GNSS has several benefits.  First, there is no need for
cooperation with the constellation owner, which eliminates a potential
coordination barrier to offering a PNT service.  Second, users may exploit LEO
SoPs without compensating the constellation owner, as has been the case with
terrestrial cellular SoPs\cite{kassas2020sops}.  Third, since SoP-based PNT is
necessarily passive, it preserves users' anonymity.  Taken together, these
three advantages are unique to SoP-based PNT and cannot be directly matched by
non-opportunistic techniques.

Despite these advantages, SoP-based LEO GNSS suffers a key limitation, which
might be termed the ``few-in-view'' problem.  With fewer than four (or, in the
case of Doppler-based PNT, eight) satellites in view, near-instantaneous
cold-start PNT with inexpensive clocks is not possible: the time to achieve a
PNT fix stretches from seconds, as with traditional GNSS, to several minutes,
as with TRANSIT, Iridium, and Globalstar \cite{j_spilker96_dop,
  lawrence2017navigation, psiaki2020navigation, mclemore2020navigation}.

One might expect LEO mega-constellations to provide greater SV coverage for
SoP-based PNT than do the relatively small Iridium and Globalstar
constellations.  However, a large fraction of mega-constellation SVs will orbit
at altitudes lower than Iridium and far lower than Globalstar, offering smaller
terrestrial service areas per
vehicle\cite{starlink2021parameters,kuiper2019parameters}.  Moreover, not all
overhead satellites may direct energy to a given user's location.  Although
early SpaceX regulatory filings indicated its Starlink mega-constellation SVs
would broadcast a quasi-omni-directional beacon signal to aid network entry, it
is not clear whether such a beacon will always be present in the system as
launched.  Recent work by Neinavaie et al. detected Doppler-trackable beacons
~\cite{neinavaie2021exploiting}, but in a contemporaneous Starlink signal
analysis the present paper's authors found that such narrow-band emissions
appeared to be absent when the downlink was busy.  Thus, beacon signals may
only be sent when the downlink is idle, rendering them intermittent or totally
unavailable once the system is more fully burdened.

Consequently, the only SoPs available from Starlink may be the broadband
signals carried in narrow spot beams from each SV toward a small number of
assigned compact service regions \cite{starlink2020redditAma}.  Significantly,
the present authors' Starlink signal analysis has revealed that each service
region is illuminated by broadband signals from at most two SVs.  Thus, areas
with no active subscribers may receive no broadband signals at all. Other
broadband mega-constellation operators will likely adopt designs similar to
Starlink's.  The net effect, at any given instant, will be a reduction in the
number of satellites actively illuminating the SoP user's location.
\ref{supp:sops} analyzes a scenario in which the global average number of
SoPs from a LEO mega-constellation is less than that of Iridium by a factor of
$\SoPCoverageRatio\times$.  This takes single-mega-constellation-based SoP LEO
GNSS from one-in-view to less-than-one-in-view, with a time to fix that will be
unacceptably long for many applications.

\subsection{Fused LEO GNSS}
Cooperation with mega-constellation operators could solve the few-in-view
problem, enabling nearly-instantaneous-time-to-fix global PNT via
traditional-GNSS-like multi-lateration.  In this paradigm, PNT becomes a
secondary service that augments the LEO mega-constellations' primary
communications mission.  Befitting its ancillary status, the PNT service ought
not require significant changes to the SVs or to the constellation's allocation
of on-orbit resources.  This paper therefore focuses on solutions which
``fuse'' the requirements of PNT into the existing capabilities of the
mega-constellation.  In fused LEO GNSS, the hardware already designed and the
spectrum already allocated for the satellites' primary broadband mission is
dual-purposed for PNT.  While this is also true of SoP LEO GNSS, fused LEO GNSS
goes further to fully exploit the broadband signal's capabilities.

To support a fused LEO GNSS service, the constellation operator arranges for
intermittent spot-beam coverage of areas where PNT users are present, providing
signals from enough satellites for receivers to produce single-epoch
stand-alone PNT solutions.  Such cooperation also has the benefit of
eliminating the duplication of effort associated with third-party tracking of
orbits and clocks for a dense constellation.

Compared to hosted-payload LEO GNSS, fused LEO GNSS sacrifices nothing in
performance while eliminating the costs of special-purpose on-orbit hardware.
In fact, where previous proposals targeted positioning precision on-par with
traditional GNSS pseudoranging (on the order of \SI{3}{m}), fused LEO GNSS can
improve on this by more than an order of magnitude \cite{iannucci2020fused}.
Moreover, it offers a significant anti-jam advantage over L-band hosted-payload
solutions in terms of tolerable signal-to-interference ratio, thus making it
attractive as a means for delivering assured PNT (A-PNT).  This advantage comes
at the cost of larger and potentially more expensive user equipment as compared
to a hosted payload solution: for maximal anti-jam performance, a fused LEO
receiver will require a phased array antenna.  But for many applications, the
user equipment, like the satellite hardware, will be dual-purposed for both
communications and PNT: the same mass-market antenna and radio connecting a
vehicle to a LEO communications network will be used for positioning at little
additional cost.

These strengths emerge from two features of fused LEO GNSS. First, the
plentiful data bandwidth present in each broadband satellite transmission burst
permits supplying users with up-to-the-instant (and therefore highly accurate)
orbit and clock products.  Such orbit and clock products need not depend on
atomic clocks onboard the SVs nor an extensive SV-observing network on the
ground. Instead, the PNT service can employ a multi-tier GNSS architecture in
which each SV's orbit and clock models are obtained via on-orbit precision
orbit determination (POD) based on an onboard traditional GNSS receiver driven
by a modest-quality clock \cite{Reid2016LeveragingCB}.  Second, unlike
traditional L-band services, commercial broadband signals in K-band and V-band
have both high signal-to-noise-ratio (SNR) and large bandwidth.  This greatly
reduces receiver noise and multipath as a source of user ranging error, even
when the ranging signal used over the communications link adopts the same
structure and spectral profile as the usual communications signals.
Furthermore, because these signals have a much shorter wavelength than
traditional GNSS, it is possible to build a highly-directional receiver phased
array for an additional \SI{30}{dB} of anti-jam performance that is compact
relative to its L-band equivalent.

PNT precision, anti-jam performance, and other constellation characteristics
are compared in Table~\ref{tab:hostedvsfused} for traditional GNSS,
hosted-payload LEO GNSS, and fused LEO GNSS.  SoP LEO GNSS is not included due
to its few-in-view problem.

For use cases in which a hemispherical antenna is preferred, such as handheld
devices, the fused SNR is not high enough to permit ephemeris and clock model
updates via the standard broadband data link. Thus, a back-up communications
link such as cellular data service would be required.  Note that certain design
elements that give fused LEO GNSS its performance advantage could be
incorporated into future hosted payload proposals.  However, this paper only
makes comparisons against published proposals.


%
%

\newcommand{\hditto}{\;\ \leaders\hrule height.5ex depth\dimexpr-.5ex+0.4pt\hfill\;\ }

\begin{table}[t]
    \centering
    \tikzset{
        cross/.pic = {\draw[thick] (-#1,-#1) -- (#1,#1); \draw[thick] (#1,-#1) -- (-#1, #1);},
        check/.pic = {\draw[thick] (-1.25*#1,-.25*#1) -- (-.5*#1,-#1) -- (1.5*#1, #1);}
    }
    \newcommand{\yes}{\tikz[anchor=base,baseline]{\path (0,.25em) pic {check=.32em};}}
    \newcommand{\no}{}
    \newcommand{\ish}{$\dagger$}
    \sisetup{range-phrase=--,range-units=single}
    \begin{tabular}{p{.3\linewidth} >{\centering}m{.1\linewidth} >{\centering}m{.12\linewidth} >{\centering}m{.12\linewidth} >{\centering\arraybackslash}m{.12\linewidth}}
        \toprule
        Characteristic           & \shortstack{Traditional \\ GNSS} & Hosted\cite{Reid2018BroadbandLEO} & \shortstack{Fused \\ (hemi RX)} & \shortstack{Fused \\ (array RX)} \\
        \midrule
        Single-epoch PNT         & \yes   & \yes   & \yes & \yes \\
        Unlimited users          & \yes   & \yes   & \yes & \yes \\
        Low Earth Orbit          & \no    & \yes   & \yes & \yes \\
        Mega-constellation       & \no    & \yes   & \yes & \yes \\
        On-orbit POD             & \no    & \yes   & \yes & \yes \\
        Non-atomic clocks        & \no    & \no    & \yes & \yes \\
        Time multiplexed         & \no    & \no    & \yes & \yes \\
        Excess bandwidth         & \no    & \no    & \yes & \yes \\
        Zero age-of-ephemeris    & \no    & \no    & \ish & \yes \\
        Highly directional       & \no    & \no    & \no  & \yes \\
        \midrule
        Localized power boost    & \$\$\$ &  \$    & \$   & \$   \\
        \midrule
        Precision \:\Centerstack{horz.\\vert.} &
            \shortstack{\SI{3.0}{m} \\ \SI{4.8}{m}} &
            \shortstack{\SI{3.0}{m} \\ \SI{4.4}{m}} &
            \shortstack{\SI{37}{cm} \\ \SI{48}{cm}} &
            \shortstack{\errorrange} \\
        Anti-jam advantage       & --- & \SI[retain-explicit-plus]{+30}{dB} & \SI[retain-explicit-plus]{+25.3}{dB} & \SI[retain-explicit-plus]{+56}{dB} \\
        \midrule
        Maturity                 & Mature & \multicolumn{3}{c}{\hditto Unproven\hditto}  \\
        Funding                  & Public & \multicolumn{3}{c}{\hditto Private\hditto} \\
        Cost to user             & Gratis & \multicolumn{3}{c}{\hditto Commercial\hditto} \\
        \bottomrule
    \end{tabular}
    \caption{Contrasting traditional GNSS, previous hosted-payload proposals,
      and fused LEO GNSS. Precise orbit determination (POD) here assumes
      onboard GNSS receivers in LEO (multi-tier GNSS).  Positioning precision
      is $95^{\rm th}$ percentile in the horizontal and vertical directions.
      Anti-jam advantage is compared to an L-band choke-ring antenna
      \cite{iannucci2020fused}.  Because K-band downlink power is tailored to
      meet power flux regulations at ground level\cite{starlinkgen2}, variable
      atmospheric absorption due to e.g. weather is assumed to be compensated by
      increased transmit power at the SV.\\
      {\em $\dagger$ If user downloads ephemeris via some other channel.}}
    \label{tab:hostedvsfused}
\end{table}


To be viable, a fused LEO GNSS service must be cost-effective for providers.
As one of its key contributions, this paper shows that providing PNT service to
every user in one service cell (e.g., for the Starlink constellation, a hexagon
of up to \cellAreaMax~\cite{starlink2019parameters})
is roughly as costly, in terms of constellation resources spent providing PNT
signals, as a single \broadbandEquivalentService\ downlink stream.  
Also, whereas broadband service expends constellation resources in proportion
to the number and activity level of subscribers, GNSS service consumes
resources in proportion to coverage area.  For this reason, in dense urban
centers where only a small fraction of potential broadband subscribers can be
accommodated and alternatives for broadband connectivity abound, a fused LEO
GNSS service could be a profitable complement to a mega-constellation's primary
broadband mission.

Indeed, it has been observed \cite{osoro2021techno} that effective subscriber
density constraints in first-generation \Ku{} broadband LEO systems could be
severe.  For this reason, population distribution statistics are invoked only
indirectly in what follows, insofar as they are needed to predict the global
distribution of downlink power expenditure onboard the SVs.


\section{Broadband LEO Concept of Operations}
This section presents a summary discussion, not of fused LEO GNSS, but of the
type of modern broadband LEO system upon which fused LEO GNSS may be built.
This digression is necessary for two reasons.  First, fused LEO GNSS is an
exercise in re-use.  To re-arrange the building blocks of a broadband LEO
system into a GNSS, one must first identify these building blocks and
understand their operation.  Second, a provider contemplating fused LEO GNSS
faces costs arising from lost opportunities for profitable broadband service.
Quantifying such opportunity costs, as will be attempted in a later section,
requires an understanding of the resource constraints that will dominate the
cost analysis.  This section summarizes relevant portions of what is known, and
lays out reasoned speculation about what is unknown, regarding the concept of
operations of broadband LEO systems.

Plans for each of the leading broadband LEO projects, including SpaceX's
Starlink, Amazon's Kuiper, and OneWeb, envision thousands to tens of thousands
of space vehicles (SVs) in orbits ranging from \SI{335}{km} to \SI{1325}{km}
altitude.  These systems are proposed to provide broadband, worldwide
connectivity to consumers via the \Ku{} and \Ka{} microwave bands between
\SIrange{10}{30}{GHz}.  Compared with existing infrastructural wireless systems
like LTE and Wi-Fi, broadband LEO systems offer greater availability in remote
locations at a lesser infrastructural investment. In densely-populated regions,
the systems are not expected to compete with cable, fiber, and 5G to serve a
large fraction of users, but would still offer competitive bandwidth and
latency to a limited user subset.

To the extent possible, this paper will remain agnostic to the details of any
particular broadband LEO system.  However, it will be helpful to refer to
concrete examples at some points in the discussion.  At such points, this paper
will refer to information gleaned from public statements and filings with the
U.S.  Federal Communications Commission (FCC) regarding SpaceX's Starlink
constellation, currently the most mature and ambitious broadband LEO contender,
with plans for for up to 42,000 LEO spacecraft.

\subsection{Architecture}
A broadband LEO SV functions as a cross between a wireless router and a
cellular base-station.  It operates according to a dynamic {\em schedule} which
allocates resources of time, space, and frequency to route packets of data among
gateways and users.  {\em Users} are paying subscribers equipped with
directional transceivers (``modems''), mounted either on static or vehicular
platforms.  {\em Gateways} are special ground sites with (potentially)
superior antennas, unobstructed skylines, and high-bandwidth, low-latency
connections to the Internet.

\subsection{Initialization}
After power-on, a user's modem attempts to enter the network by searching for
information that will allow it to connect to an SV: orbital parameters,
positioning, timing, and antenna orientation.  Any parameters that it cannot
obtain or recall, it must determine by a guess-and-check strategy.  The user
modem tunes its phased-array antenna to point in certain directions where an SV
might be found, and then either listens for a signal on a particular downlink
frequency, or transmits a connection request on a particular uplink frequency.
Until it receives a signal or a response, it keeps trying with other directions
and frequencies.  Once a connection is established, the user modem and network
exchange authentication tokens.


\subsection{Steady-state operation}
After initialization, operation likely proceeds as in Wi-Fi or cellular
communications with central scheduling: the user modem may notify the SV,
within certain windows of time and frequency, that it wishes to uplink a packet
of e.g., Dynamic Host Configuration Protocol (DHCP) or Internet Protocol (IP)
data.  At its discretion, the SV grants the user modem an uplink time-slot.
The modem listens for such a scheduling directive, waits for the appropriate
moment, and then encodes and transmits the packet.  Conversely, when a packet
addressed to a user arrives at the SV from a gateway, the SV queues the packet
for downlink transmission.  The user modem monitors the downlink channel for
messages.

As with Wi-Fi or any other wireless Internet medium, the broadband LEO
architecture need not make any special affordance for higher-level abstractions
like data fragmentation and re-assembly, reliable and in-order delivery,
Internet sockets, or applications like the Web: these features are provided by
IP and communication layers built atop it, like the Transmission Control
Protocol (TCP) and the Hypertext Transport Protocol (HTTP), in accordance with
the so-called end-to-end principle.

\subsection{Error correction and re-transmission}
Each transmitted packet 
is encoded using error detection (e.g., cyclic redundancy check) and error
correction (e.g., Turbo codes) mechanisms.  If a packet is decoded
successfully, the receiver (user modem or SV) sends back an acknowledgment
(ACK).  If a packet fails to decode, the receiver sends back a negative
acknowledgment (NAK).  A packet which has been NAK'd may be scheduled for
another transmission attempt using automated repeat request (ARQ) or hybrid ARQ
(H-ARQ).


Fused LEO GNSS will involve tasking the downlink transmitter to send additional
PNT-specific signals.  Because the SV knows the timing of these additional
signals in advance, it must schedule around them in provisioning data service:
uplink packets and their re-transmissions must be scheduled so that an ACK need
not be sent down while the SV transmitter is busy, and downlink packets and
their re-transmissions must be scheduled so that no data transmission is
required during transmission of a PNT signal.  Section
\ref{ss:schedulingconstraints} offers further details on such constraints.

All broadband LEO systems will likely employ frequency-division duplexing
(FDD), whereby the uplink and downlink operations are frequency-disjoint.  The
alternative, time-division duplexing (TDD), is wasteful for long-distance
channels: avoiding simultaneous reception and transmission at both ends
requires that the channel frequently sit idle.  As with FDD LTE, the user
modems will likely be frequency-division half-duplex (one frequency at a time)
to keep modem cost low, whereas the SVs, like cellular stations, will support
full FDD.  This implies that an SV need not wait for an ACK to a downlink
packet before it begins transmitting another packet.  As will be discussed
later, this property is significant for fused LEO GNSS.


\subsection{Multiplexing}
General wireless principles offer a number of options for partitioning a shared
downlink channel among multiple users, or among distinct data streams,
including code-, time\nobreakdash-, frequency-, and space-division multiplexing
(C/T/F/SDM).  Typically, several techniques are employed to achieve high
spectral efficiency at modest computational cost.  K-band broadband LEO systems
operate on a dynamic, wide-band, dispersive channel\cite{cid2015wideband}.
These are the same parameters that characterize the radio-layer design of
existing mobile wireless networks like LTE and Wi-Fi, and hence one may expect
similar solutions to apply.

LTE and Wi-Fi use orthogonal frequency-division multiplexing (OFDM) to
eliminate dispersive inter-symbol interference due to, e.g., multipath
scattering, and SDM in the form of beamforming to boost signal strength.  This
paper presumes that broadband LEO systems will also apply OFDM, acknowledging
its preeminence in modern wireless technologies. OFDM's spectrally-flat signals
are not theoretically ideal for ranging, but when wide enough offer excellent
precision \cite{iannucci2020fused}.  Beamforming, discussed further in
\S\ref{ss:beamforming}, is critical to close link budgets and to avoid causing
interference.  Apropos of fused LEO GNSS, beamforming implies that
signals-of-opportunity from broadband LEO SVs cannot be expected to be strong
enough for navigation unless a ``spot'' beam is directed towards the listener's
geographical vicinity.

The third form of multiplexing likely to be in use---and relevant for fused LEO
GNSS---is time division.  In FDD LTE, the channel is licensed to the mobile
network operator, and is provisioned for a single transmitter in a given
spatial region.  Time division is used for sharing resources between users,
with time being divided into frames, subframes, and OFDM symbols, and users
being scheduled downlink opportunities in the time-frequency plane.  In
broadband LEO, one might expect the same to apply.  However, continuous
operation of the SV transmitter regardless of load would waste a
significant amount of power: wide-band OFDM signals have a high peak-to-average
ratio (PAR) and require a high degree of linearity, and PAR tends to drive
power requirements for all but the most sophisticated linear amplifiers.  Power
may be a significant constraint for the operations of broadband LEO systems
(\S\ref{sec:power-management}).  This paper therefore assumes that broadband
LEO transmitters operate in ``burst mode,'' i.e., with intermittent
transmissions.  Such a mode aligns radio power consumption with load.

%

\subsection{Beamforming}
\label{ss:beamforming}
LEO signals are fundamentally visible to a much smaller portion of the Earth's
surface than are traditional GNSS signals.  Furthermore, the requirement of a
broadband system for simultaneous high SNR and bandwidth is possible only by
focusing an SV's transmit power into a narrow beam targeted toward a relatively
small ground service region.  Each SV therefore must have
independently-steerable directional antennas for transmit and for receive
operations.  Each Starlink SV, for instance, will support \Nbeams\ downlink
beams for user data service~\cite{starlink2019parameters}.

\begin{figure}[htp]
    \centering
    \includegraphics[width=\columnwidth, trim=5 10 12 0, clip]{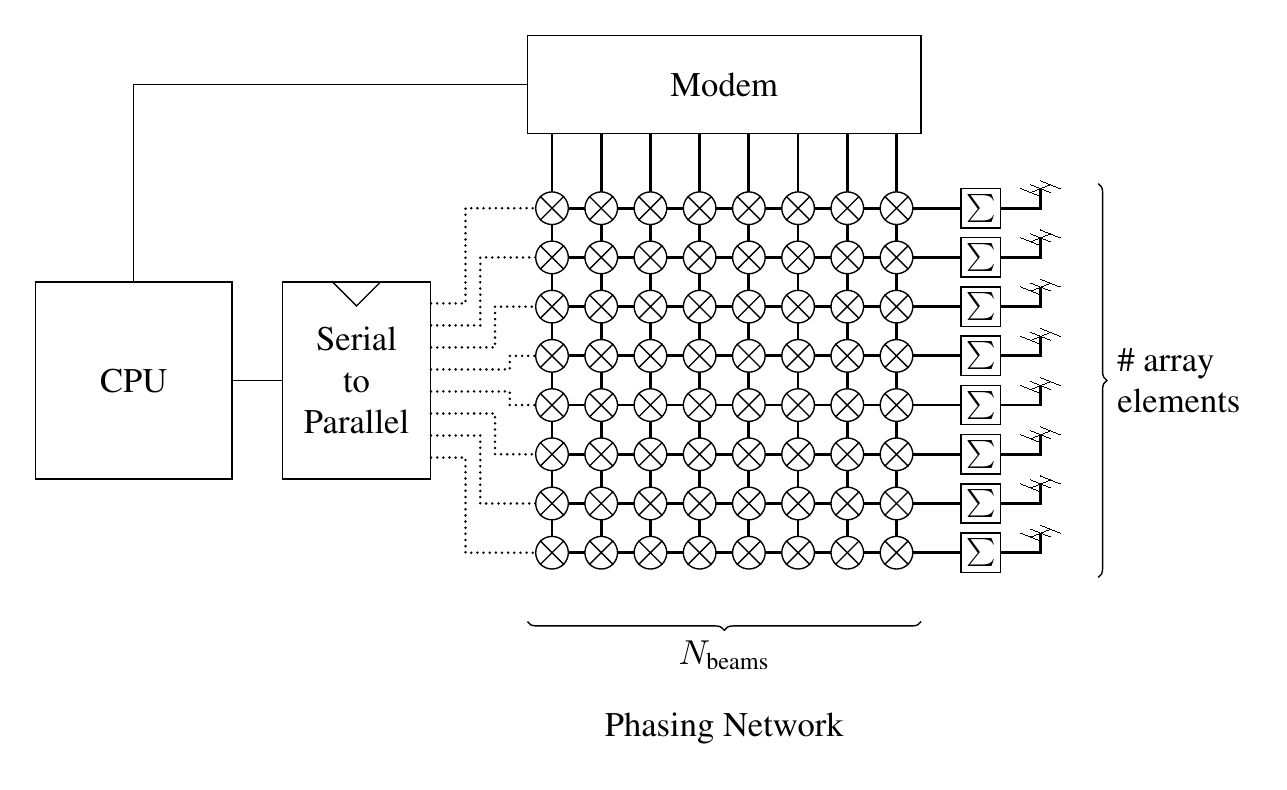}
    \caption{An electronically steered phased array for transmission.
      Amplifiers not shown.  Columns are beams; rows are array elements.
      Dotted lines carry coefficients; thick solid lines carry radio frequency
      (RF) signals.  For reception, the matrix of phasing elements and
      accumulators is transposed: the accumulators move from the right
      (antenna) lines to the top (modem) lines.}
    \label{fig:serialtoparallel}
\end{figure}

Because each LEO SV is overhead only as viewed from a relatively small region
at a time, a large number of SVs are needed to provide continuous global
service.  To mitigate launch costs, each SV must be (relatively) light and
compact.  This configuration strongly favors flat, electronically-steered
two-dimensional phased array antennas.  Such antennas require costly {\em
  phasing networks}.  These are (analog or digital) circuits which form
programmable combinations of one or more transmitted signals (``beams'') for
each one of a large number of radiators (``array elements'').  A schematic
representation is shown in Fig.~\ref{fig:serialtoparallel}.  The contribution
of each beam to each array element is linear, and consists of a delay and/or
scaling (from now on, ``phasing'').  Mathematically, this operation is a
complex matrix multiply for each frequency, but such a cavalier description
belies significant implementation challenges.

For this paper, what is relevant is the procedure for ``steering'' the array.
The phasing elements require multi-bit coefficients for configuration.  During
steering, the SV's central processing unit (CPU) must compute (or retrieve from a
look-up table) a new set of coefficients for the desired beam pattern.  These
coefficients must then be transferred to the phasing network.  To free up the
CPU's I/O resources for other tasks, direct control over the phasing elements
would typically be delegated to an external multiplexer, shown in
Fig.~\ref{fig:serialtoparallel} as a serial-to-parallel converter.  (This
example architecture is not the only way to design such a circuit; it serves
only to facilitate the parameter definitions needed for this paper.)  Over some
interval $T_\text{set-up}$, the CPU loads coefficients into this multiplexer.
Then, on a signal from the CPU, the updated coefficients are simultaneously
imposed on the phasing network.  Such an arrangement is favorable because it
minimizes the ``downtime'' during steering in which the phasing network is in an
indeterminate state and so cannot be relied upon to produce a valid beam
pattern.  The downtime could be made as short as a few times the light-crossing
time of the phased array: some few nanoseconds.  The downtime will be denoted by
$T_\text{switch}$ in what follows.



%

Note that $1/T_\text{set-up}$ is an upper limit on the rate at which the array
may be re-steered.  $T_\text{switch}$ may be expected to be shorter than
$T_\text{set-up}$, so while it also imposes an upper limit, it is a less
stringent one.  Also, for reasons of cost, it is reasonable to assume that user
modems will only be capable of transmitting or receiving a single beam at a
time.


\subsection{Channels}
The number of disjoint frequency channels available is a function of spectrum
licensing and desired system bandwidth to a single user.  Public filings for
Starlink indicate that the K-band spectrum in the ranges
\SIrange[range-phrase=--]{10.7}{12.7}{GHz},
\SIrange[range-phrase=--]{17.8}{18.6}{GHz},
\SIrange[range-phrase=--]{18.8}{19.3}{GHz}, and
\SIrange[range-phrase=--]{19.7}{20.2}{GHz}
will be broken into $\Nchannels\times\W$ downlink channels.  This paper does not
assume that all channels are available on all beams, nor that all available
channels may be transmitted simultaneously by any one beam.  Instead, it assumes
that there is some number $N_\text{bc}$ of ``beam-channels'' that represents the
greatest number of simultaneous transmissions from the SV.  For Starlink, this
number is at most \Nbc, though it may be less.
When concrete values are required in what follows, this paper will assume
$N_\text{beams}=\Nbeams$, $N_\text{channels}=\Nchannels$, and
$N_\text{bc}=\Nbc$.

\subsection{Flux}
\label{ss:flux}
One consideration particular to first-generation broadband LEO proposals has
been driven by the opening of spectrum in the K band.  Operators can access
this spectrum only by adhering to certain limits on the flux of RF energy at
the surface of the Earth intended to prevent interference with terrestrial
K-band services.  For instance, Starlink has declared that the system will
generate no more than \PFD\ in the \SIrange{10.7}{12.7}{GHz} band as observed
at ground level at elevation angles above \SI{25}{\degree}.

It has been revealed in public
filings for Starlink that a single beam from a single broadband LEO SV is
powerful enough and directional enough to saturate this flux limit.  While a
shrewd operator might wish to increase system capacity in densely-populated
urban regions by focusing several beams from several SVs onto one location,
taking advantage of user antenna directionality to boost spatial re-use, this
would not avail: each beam would have to operate with either reduced power or
reduced duty cycle to avoid exceeding the flux limit.


Note, however, that the flux limits are per-MHz, so an operator might be
permitted to focus multiple beams on the same location, provided these are on
disjoint channels.  This solution might cause other problems if it resulted in
all available channels being occupied serving a dense region, leaving none
available to serve a nearby sparse region due to the possibility of
same-channel interference.

\subsection{Cellular Provisioning}
\begin{figure}[htp]
    \centering
    \includegraphics[width=\columnwidth]{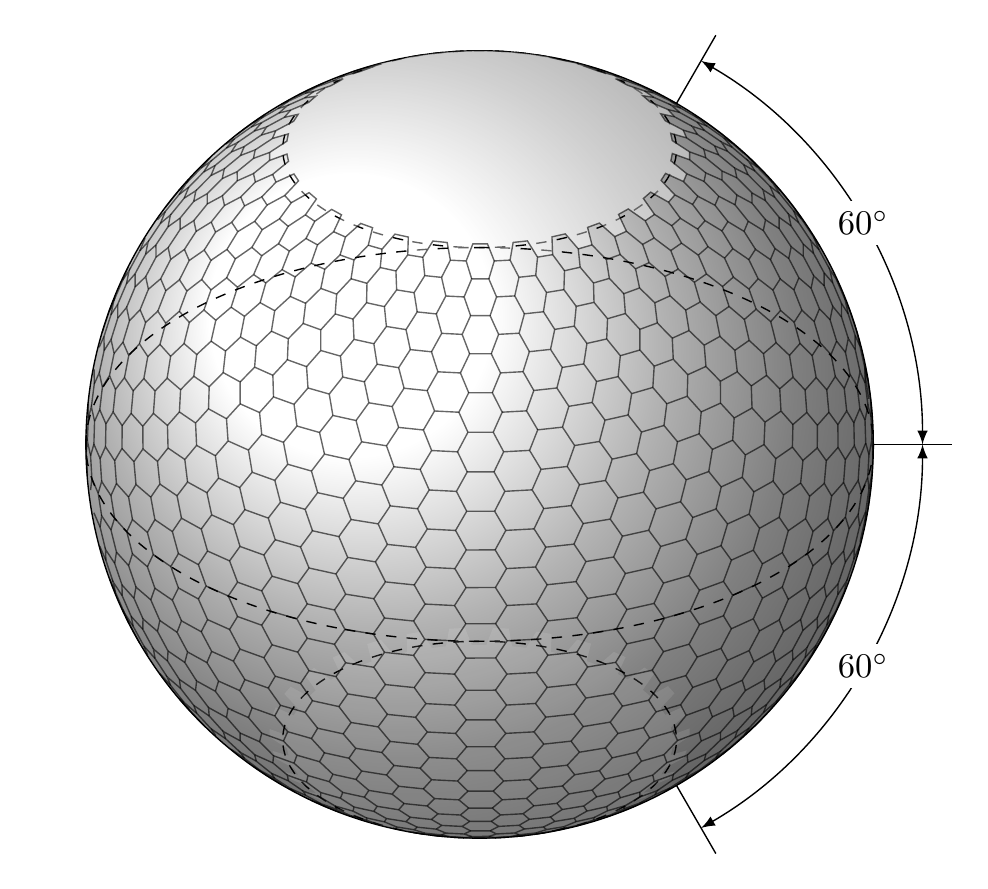}
    \caption{Schematic geometry for geographical provisioning of broadband
      service for a LEO system covering the range of $\pm\SI{60}{\degree}$
      latitude.  Real-world cells would be smaller, with diameters $D$
      measuring in the tens of kilometers.  The area of the service region is
      $4\pi R_\text{earth}^2 \sin \SI{60}{\degree}$, and the area of an
      individual cell is $\frac{3\sqrt3}{8} D^2$, ignoring north-to-south
      variations. For Starlink, the number of cells is approximately equal to
      the surface area of the Earth between $\pm\serviceLatMax$ latitude
      divided by the area of a hexagon of diameter $D = \D$: this gives a
      number on the order of \SI{\Ncells}{}.}
    \label{fig:cells}
\end{figure}

To avoid same-channel interference between users served by different SVs,
broadband LEO operators will establish regions of exclusion within which a
channel may not be re-used.  The simplest way to do this would be to provision
broadband LEO service into a grid of hexagonal cells on the surface of the
Earth (Fig.~\ref{fig:cells}).  Rather than assigning individual users to
individual SVs, users are assigned to a local cell, and the cell is assigned to
an SV.  Provided that SV antennas are suitably directional, a channel used in
one cell may be re-used in other cells so long as these cells are not immediate
neighbors.  Provided that signals to each individual user are suitably
multiplexed in time and/or frequency, a channel may serve a large number of
users within one cell, just as in LTE.

One refinement would be to rely on user antenna directionality to avoid
same-channel interference in adjacent cells while permitting additional spatial
re-use.  This might, however, run afoul of the flux limits, and would have to
be done with care to maintain a sufficient angular separation between the
assigned SVs.  Even in the best case, the need for angular separation would
tend to push the typical SV assignments for each cell away from the zenith,
potentially reducing availability for users with less-than-totally-clear skies.

\subsection{Cellular Scheduling}
For reasons of latency and efficiency, it makes sense to delegate, insofar as
possible, the global network coordination and packet scheduling problem to the
SVs.  Setting aside for the moment the question of how SVs know which cells to
serve on which beams and frequencies, each SV may independently make
millisecond-level decisions about the scheduling of uplink and downlink
packets; these decisions have no impact on other cells.  The alternative would
be for SVs to consult with their gateways, or even with a central server,
before a user's request for data could be satisfied, which would needlessly
incur several milliseconds of additional latency and tie up gateway bandwidth.

On the other hand, it is not easy to fully decentralize the global network
coordination problem.  One could imagine designing a fixed, pre-arranged
schedule of assignments of SVs to service cells, but this scheme fails on
several counts: (1) As mega-constellations are gradually deployed, the schedule
must be constantly adjusted to account for unfilled orbits.  (2) Dense
deployments like Starlink are expected to be spread over a number of ``shells''
at different orbital altitudes, making their global configuration aperiodic.
Even if orbital periods are commensurate, a mega-constellation will necessarily
be in constant flux as SVs are added or replaced.  (3) A pre-arranged schedule
fails to accommodate dynamic variations in the health, level of battery charge,
and access to gateway bandwidth of individual SVs.

It would be straightforward to periodically re-compute, at a central location,
the assignment of SV beams to cells and gateways to SVs, and for commands
effecting this schedule to be uplinked to the SVs via the gateways.  Whether
operating at initial operational capability or full capability, tasking each SV
beam to serve many cells or just one, such central coordination would give the
operator much-needed flexibility.

How often must this computation take place?  The orbital motion of LEO SVs
across the user's sky plays out over a period of several minutes, so that
during the interval from when an SV becomes the ``most suitable'' to serve a
given cell to the time that it is no longer the most suitable, hundreds of
thousands of packets may be sent.  Assignments of gateways to SVs will evolve
over similar time-scales.

Downlink scheduling plays a critical role in costing out the fused LEO concept,
which involves additional steering costs and additional transmissions.  For
systems like Starlink that use frequency-division duplexing, with uplink
channels disjoint from downlink channels, the uplink schedule is expected to be
largely unaffected by fused LEO GNSS.

Will broadband LEO SVs be able to steer their (possibly paired) transmit and
receive beams separately and use them simultaneously?  Public filings do not
make this clear. However, if the differences in wavelength between the uplink
and downlink channels are significant, as is the case for Starlink, it seems
reasonable to assume that a separate set of phasing coefficients would be
required for each, if not completely separate antenna hardware, implying
independently-steerable uplink and downlink antennas at the SV.

\subsection{Power Management}
\label{sec:power-management}
Broadband LEO constellations can be expected to operate with highly efficient
power management strategies.  Their SVs' solar arrays will be sized just large
enough to meet some low multiple of the expected average energy demand per
orbit at end of life, and their batteries will be sized just large enough to
meet some fraction of peak regional demand, and to sustain operations during
eclipse \cite{mostacciuolo2018modeling}.  Regional or global scheduling
algorithms will optimize energy usage across the constellation by tasking each
active SV to nearly exhaust its collected energy per orbit.

This is not to say that SVs will operate continuously at or near their maximum
load power.  Rather, they will be commanded to enter a ``deep sleep''
minimum-power state during a portion of their orbit while nearby SVs carry the
burden of providing uninterrupted service. Such duty cycling is rational
because, just as for smartphones \cite{perez2016power}, an SV's useful work is
a nonlinear function of its expended energy.  Greater efficiency obtains when a
subset of SVs operates near maximum power and the complement operates at
minimal power than when each SV is active but not fully tasked.  One may expect
that over oceans a majority of SVs will be idled whereas over some populated
areas all will be fully tasked.

The constellation enters a scarce energy regime when the constellation-wide
energy collected per orbit is inadequate to support the demanded communications
operations.  One might expect the constellation to be designed never to enter
this regime, given that opportunities for profitable exchange of data are
lost. But as with electricity provision \cite{faruqui2017arcturus} and
terrestrial broadband provision \cite{sen2012incentivizing}, it is wasteful to
design a network for peak demand if the peak-to-average demand ratio is high.
LEO broadband providers are likely to accommodate congestion just as their
terrestrial counterparts do, namely, by throttling data speeds or by
implementing time-dependent pricing \cite{sen2012incentivizing}.

\subsection{Visibility}
For K-band operation in particular, the ITU Radio Regulations require broadband
LEO operators to avoid SV-to-user lines-of-sight that pass too close either to
geostationary orbit or to the horizon.  SV beams cannot be assigned to serve
cells for which the line-of-sight falls into one of these ``exclusion masks.''
The two types of masks reduce the number of available SV beams at low and
high latitudes, respectively.



\section{Fused LEO GNSS Concept of Operations}
\label{ss:rangingfromleo}
To achieve latency and performance competitive with traditional GNSS, fused LEO
GNSS will need to employ single-epoch code-phase pseudoranging using bursts
transmitted from four or more SVs.  This will require changes to broadband LEO
operators' coordination and scheduling algorithms: multiple SVs must provide
time-multiplexed ranging signals to each cell to support pseudorange-based PNT.
These broadcast ranging signals are an addition to the broadband LEO system,
but they share the same modulation and encoding as the data service.

\subsection{Obstacles}
A comprehensive proposal for fused LEO GNSS must address certain key obstacles:
Broadband service in a given cell is expected to be provided by a single SV for
minutes at a time.  Multi-SV-to-cell pseudoranging will require changes to both
the central scheduler that matches SV beams with cells, and to the onboard
schedulers that must avoid collisions with cross-cell ranging signals, i.e.,
those from other SVs (\S\ref{ss:schedulingconstraints}).
The downlink transmitters are not designed to produce traditional ranging
waveforms (\S\ref{ss:ranging-with-data-bursts}).
The broadband LEO SVs' positions and clock offsets may not be known to high
enough accuracy either to the broadband system or to user modems, so a precision
orbit and clock determination capability is needed (\S\ref{ss:pod}).
The onboard clocks are likely not stable enough for nanosecond-accurate
forecasting beyond a minute, so a ``zero age-of-ephemeris'' solution is
required (\S\ref{ss:zaoe}).
%
%
Finally, the SVs have limited power and limited resources for downlink
scheduling, array steering, and command-and-control signaling.  Any resources
diverted from the constellation's primary communications mission must be paid
for by PNT customers (\S\ref{ss:cost}).



\subsection{System Overview}
\begin{figure}[htp]
    \centering
    \includegraphics[width=\columnwidth]{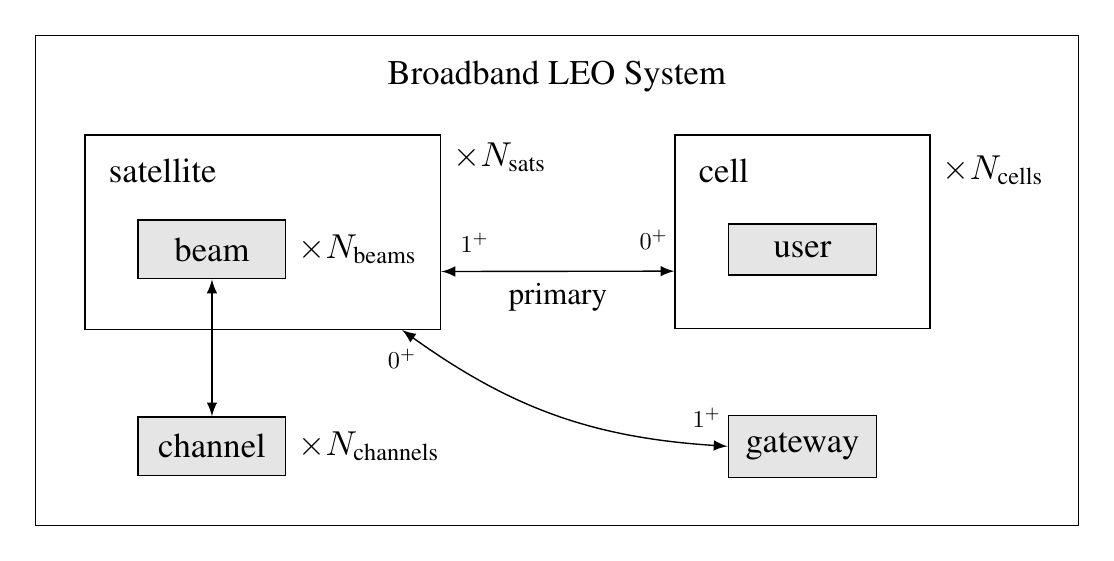}
    \caption{Entities (boxes) and their relationships (arrows) in a
    broadband LEO system.  Numbers by arrowheads indicate the cardinality of a
    relationship: for instance, the number of cells for which an SV is
    primary is zero or more, while the number of SVs which are primary
    for a cell is at least one.}
    \label{fig:commsystem}
\end{figure}

\begin{figure}[htp]
    \centering
    \includegraphics[width=\columnwidth]{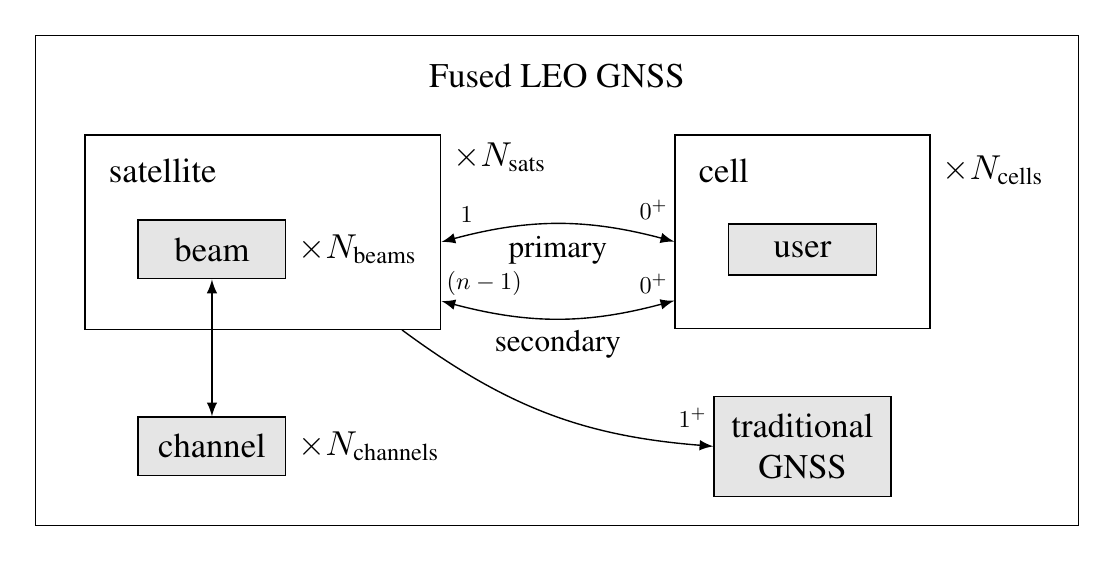}
    \caption{As Fig. \ref{fig:commsystem} but for fused LEO GNSS, with $n$
    denoting the number of ranging signals provided to users in each cell.
    Operation without traditional GNSS might be desirable but remains future
    work.}
    \label{fig:pntsystem}
\end{figure}

The underlying broadband LEO system may be depicted as in
Fig.~\ref{fig:commsystem} in terms of the relationships between SVs, beams,
cells, and users.  The fused LEO GNSS concept in Fig.~\ref{fig:pntsystem}
mirrors this structure with two alterations.  First, SV beams may receive
additional ``secondary'' cell assignments.  Each beam will broadcast periodic
ranging signals to all of its assigned cells, primary and secondary, but it
will only provide broadband service to the cells for which it is primary.
Second, whereas broadband connectivity is tied back to a gateway, PNT is tied
back to traditional GNSS signals, as observed via GNSS receivers on the SVs. In
this way, LEO provides a second ``tier'' of PNT service, with traditional GNSS
serving as the first tier (\S\ref{ss:pod}).  Note that, although dependent on
traditional GNSS, such multi-tiered fused LEO GNSS is well-protected from
terrestrial GNSS interference sources: L-band signal spreading loss to LEO is
more than 145 dB\cite{murrian2021leo}.


The total number of assignments (primary or secondary) for each served cell is
equivalent to the number of signals provided for pseudoranging, denoted $n$.
To fully constrain the three-plus-one dimensional PNT solution, $n\ge4$.  When
a concrete baseline value is required, $n=5$ will be assumed in what follows.
For a customer willing to pay a high price for ultimate performance, $n$ may be
as large as the full number of SVs in view, circa 40 \cite{iannucci2020fused}.

\begin{figure}[tp]
    \centering
    \includegraphics{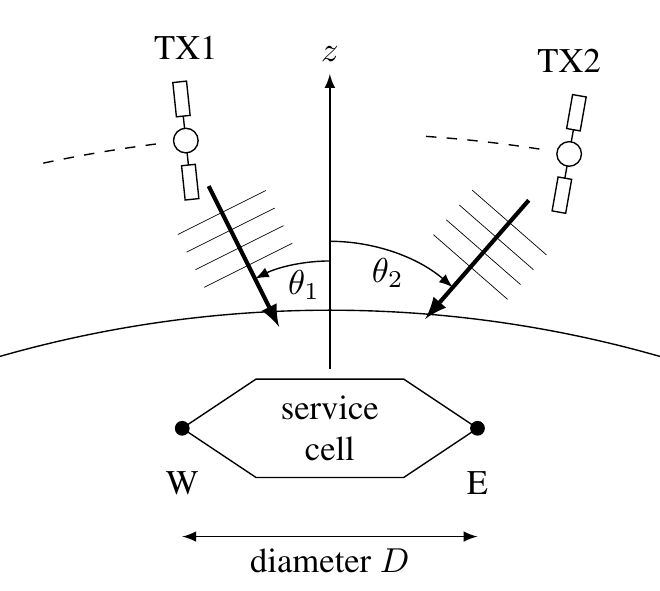}
    \caption{Two of the $n$ SVs providing ranging signals to a given cell from
      different points in the sky.  At least $n = 4$ signals from distinct
      directions are needed to solve for $x$,$y$,$z$, and time.  The
      west-to-east axis of motion shown here is arbitrary, but consistent with
      Fig.~\ref{fig:space2time}.  Angles $\theta_1$ and $\theta_2$ are the
	  zenith angles corresponding to TX1 and TX2, as observed from e.g. the
	  center of the cell.  Variation of these angles across the cell is not
	  shown.  For a satellite at altitude $h$, the variation will be on the
	  order of $D/h\approx\SI{3}{\degree}$.}
    \label{fig:wavefront_geometry}
\end{figure}

Two SVs in a ranging scenario are depicted in
Fig.~\ref{fig:wavefront_geometry}.  Users in the service cell require a
diversity of directions-of-arrival to obtain a robust positioning and timing
solution.  To maximize geometric diversity with $n=5$, the operator could
assign to a cell those SVs near the vertices of a square pyramid; e.g., the
northernmost, easternmost, southernmost, westernmost, and zenith-most SVs in
view among those that are not in an exclusion mask \cite{teng2016closed}.  (The
pyramid need not align with the cardinal directions.)

\subsection{Pseudoranging Service}
\label{ss:ranging-with-data-bursts}
Ideally, one uses a ranging signal for ranging, and a communications signal for
communications.  For fused LEO GNSS, however, it is greatly disfavored to
change the function of the SV transmitter, both because modulations tend to be
fixed in hardware, and because a high-bandwidth side-channel for orbit and
clock data embedded in the ranging burst is highly valuable.  Accordingly,
fused LEO GNSS adopts the unmodified communications waveform for both data and
ranging.  There is a small degradation in possible ranging precision arising
from this compromise, but it is dominated by other sources of error.

Service consists of a series of ranging bursts, each modulated just like
ordinary broadband data, with three exceptions.  First, the user modem must be
able to regenerate the transmitted waveform and perform cross-correlation, so
the data bits encoded and modulated to form the ranging burst must be largely
known in advance to the modem.  This cross-correlation results in a code-phase
time-of-arrival measurement, which may then be compared with the nominal
time-of-departure of the burst to form a pseudorange measurement.  Second, the
burst is very short.  A duration of $T_\text{burst}=\Tburst$ is more than
adequate: the contribution to ranging error due to finite burst duration is far
less than other sources of error\cite{iannucci2020fused}.  For this reason, it
is also not a problem to set aside a portion of the ranging burst to contain
data not known in advance to the receiver, such as up-to-date clock and orbit
ephemerides.  This portion of the burst is ignored during correlation.  Because
the entire clock and orbit ephemeris fits into a small fraction of a single
ranging burst---which might easily accommodate tens of kilobits of data---user
pseudoranges need never be based on stale or forecast data.  This ``zero
age-of-ephemeris'' eliminates the need for atomic clocks on the LEO SVs. Third,
the burst is not acknowledged by one ground receiver (i.e., unicast), but
instead is broadcast to all receivers in the cell.
\label{ss:zaoe}

One significant challenge here is that, as noted in\cite{cid2015wideband}, the
K-band channel is dispersive, with a worst-case coherence bandwidth of only
\SI{3}{MHz}.  A na\"ive cross-correlation on a \W-wide signal would not produce
a peak with width approaching $1/(\W)$; instead, in a worst case, a broad,
incoherent peak would be expected with width on the order of $1/(\SI{3}{MHz})$.
The dispersion on the channel may be decomposed into factors: (1) the frequency
response of the transmit filters, amplifiers, and phased array; (2) the
frequency response of the atmospheric channel; (3) the aggregate effect of
multipath scattering; and (4) the frequency response of the receive array,
filters, and amplifiers.  Each of these can be managed in fused LEO GNSS:
atmospheric dispersion at K-band is negligible for the bandwidths envisioned
(Hobiger et al.~\cite{hobiger2013correction} report sub-millimeter delay
sensitivity to dry air pressure, water vapor, and surface air temperature for a
\SI{200}{MHz}-wide \Ku{}-band signal), non-line-of-sight multipath effects will
be suppressed by the directionality of the receive phased array, and the
transmit and receive frequency responses may be estimated using the training
preamble that is in any event required for OFDM.

\subsection{Global Scheduling}
\label{ss:schedulingconstraints}
This paper draws a distinction between two objects which might be termed
``schedules.''  {\em Global schedules} are computed centrally and in advance by
the broadband LEO provider, and consist of assignments of SV beams to provide
service to certain cells at certain times.  {\em Local schedules} are computed
on-the-fly on each SV as packets arrive and are dispatched.  Global schedules
may be valid for multiple minutes, and can be forecast due to orbital
predictability.  Local schedules cannot be forecast due to the unpredictable
timing of packets.

\begin{figure}[tp]
    \centering
    \includegraphics[width=\columnwidth, trim=0 0 0 .5cm]{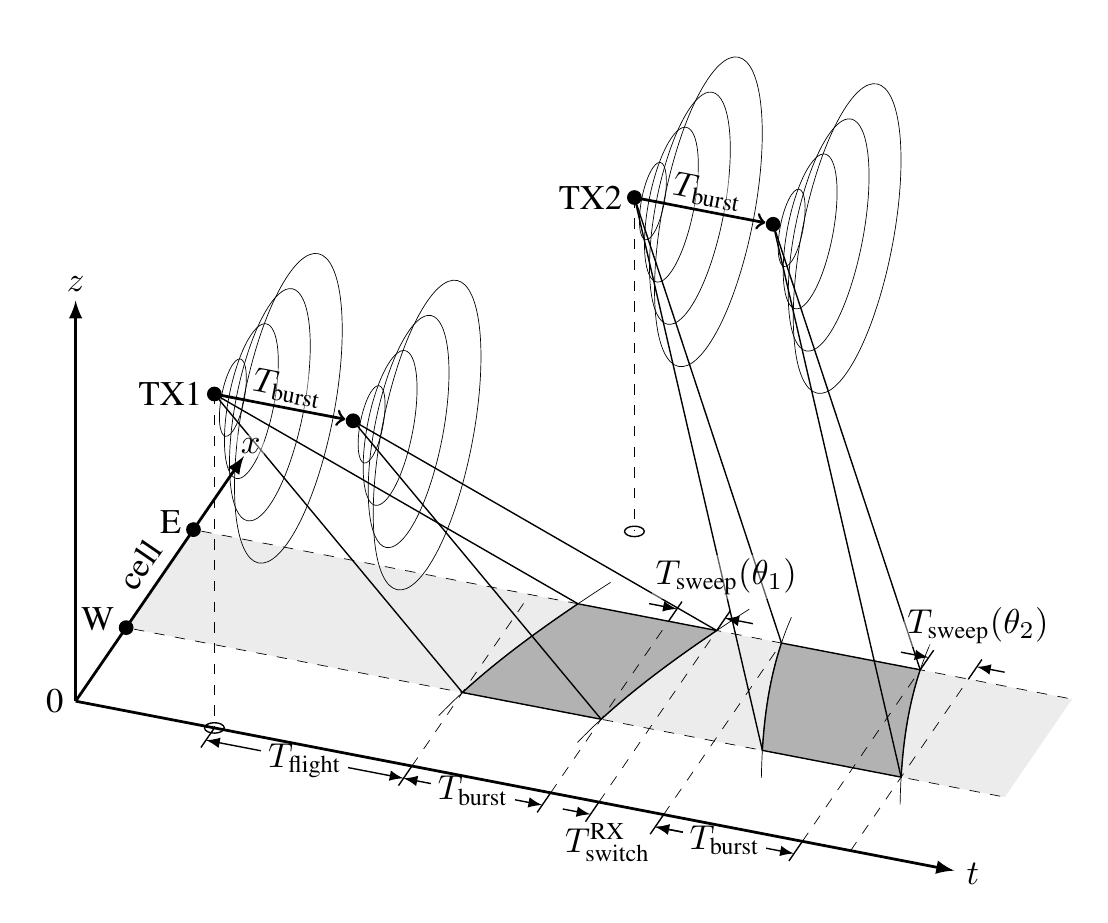}
    \caption{Satellite-to-ground bursts shown in two axes of space (east-west
    and up-down) and one of time.  The bursts begin at space-time points TX1 and
    TX2, and each continues for a duration of $T_\text{burst}$.  Signals radiate
    outward in a light-cone, intersecting the ground at $z=0$ in a segment of a
    hyperbola.  The interval until the signal first reaches the ground somewhere
    in the service cell is $T_\text{flight}$.  The interval from that time until
    the signal has reached the entire cell is $T_\text{sweep}$.
    }
    \label{fig:space2time}
\end{figure}

The global scheduler's role in broadband LEO is to compute a conflict-free
assignment of SV beams to cells: the ``primary'' assignments in the parlance of
fused LEO GNSS.  These assignments respect power, visibility, and exclusion
mask constraints.  In fused LEO GNSS, the global scheduler must additionally
solve a system of constraints on the inter-departure and inter-arrival times of
ranging bursts, particularly those ``secondary'' ranging bursts directed to a
cell from a non-primary SV.  Care is required in scheduling these cross-cell
ranging bursts so as to avoid collisions on the ground.  In
Fig.~\ref{fig:space2time}, bursts TX1 and TX2 from two different SVs arrive in
the same cell.  The wavefronts of each burst sweep across the cell in different
directions, consistent with the geometry of Fig.~\ref{fig:wavefront_geometry}.
If a user at the eastern edge of the cell is to decode both bursts (for
instance, if the first burst is data addressed to this user and the second
burst is a ranging broadcast), then an interval $T_\text{switch}^\text{RX}$
must be allowed between the end of the first burst and the arrival of the
second for the user's modem to re-tune its antenna.

\subsubsection{Feasibility}
\label{ss:schedulefeasibility}
\begin{figure}[tp]
    \centering
    \includegraphics[width=\columnwidth, trim=0 0 0 .4cm]{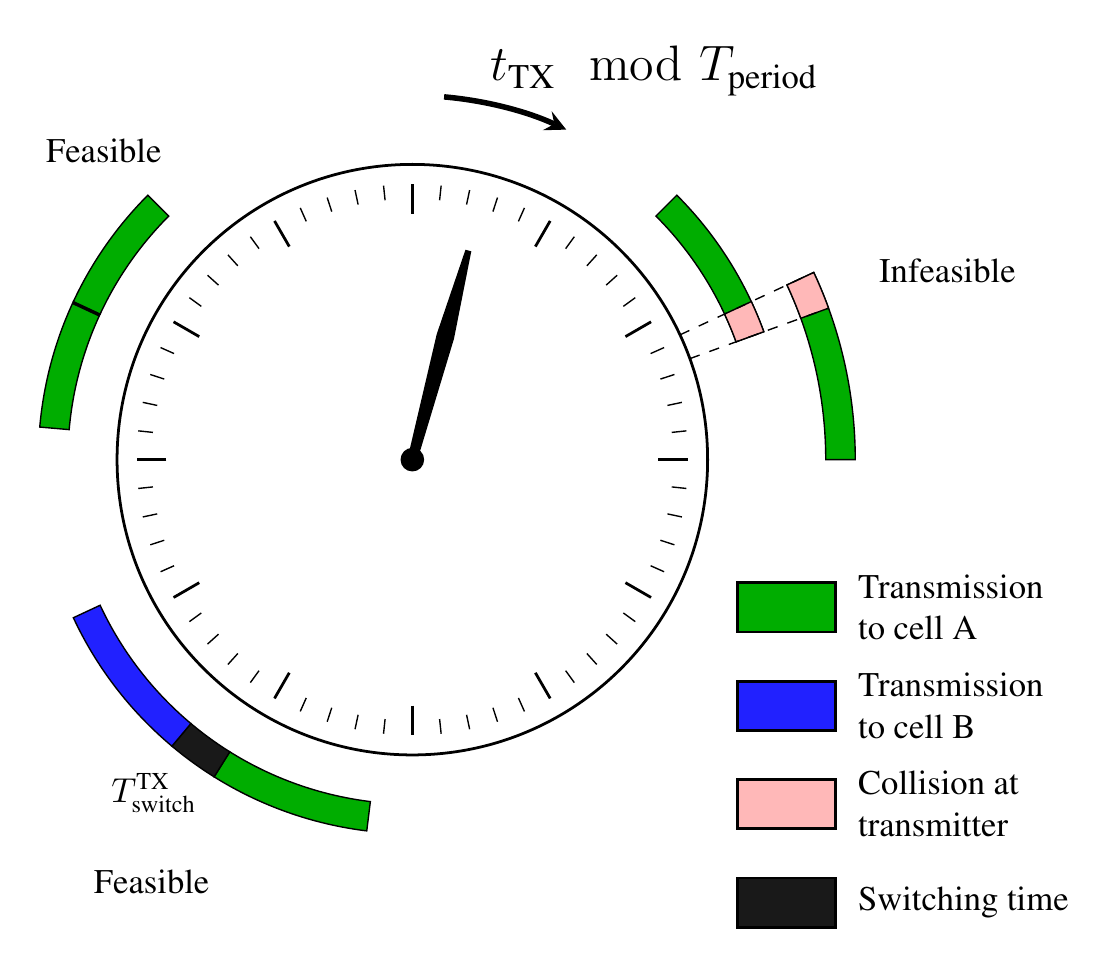}
    \caption{Scheduled events at each transmitter (i.e., SV beam) must satisfy
      feasibility constraints: bursts may not overlap, and bursts to different
      cells must be separated by at least $T_\text{switch}^\text{TX}$.}
    \label{fig:txclock}
\end{figure}

\begin{figure}[tp]
    \centering
    \includegraphics[width=\columnwidth, trim=0 0 0 .4cm]{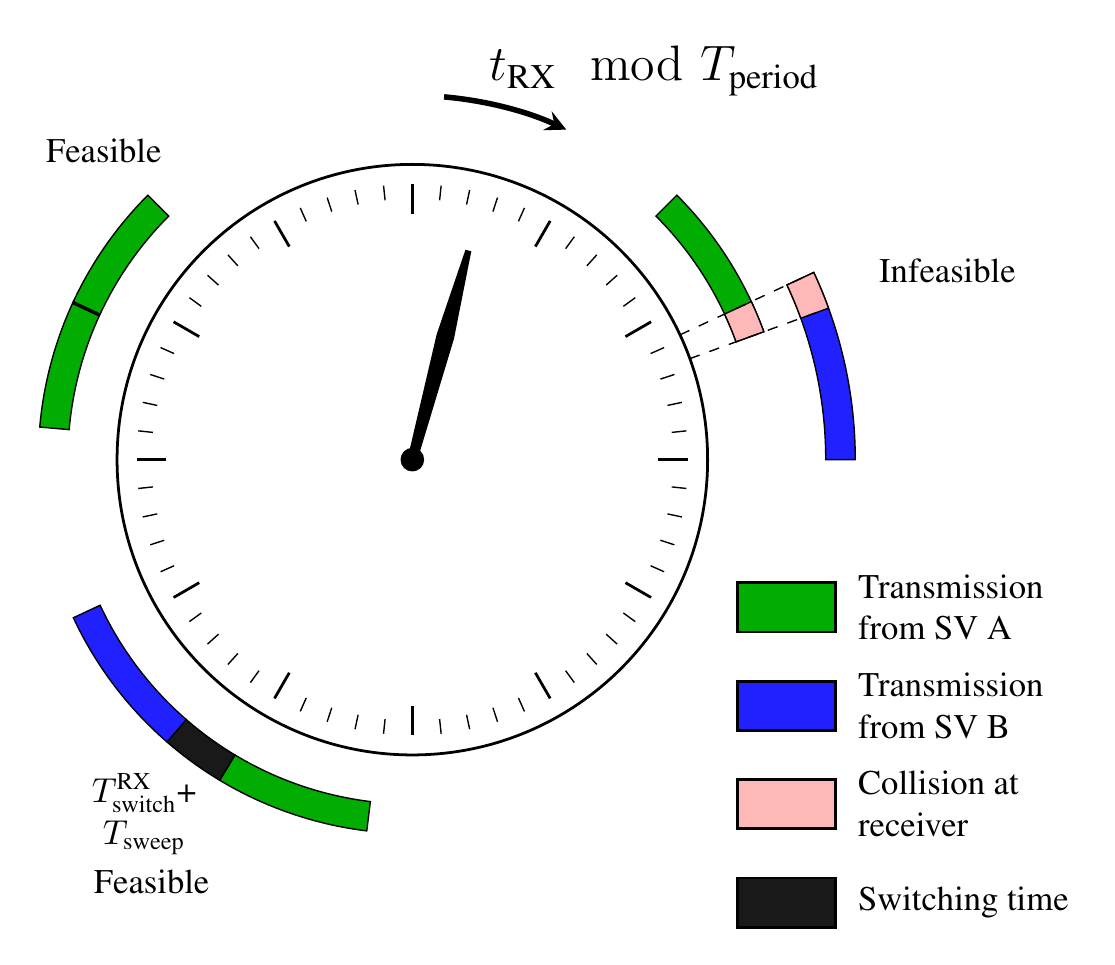}
    \caption{Scheduled events at the receiver (i.e., user modem) must satisfy
      feasibility constraints: bursts may not overlap, and bursts from
      different SVs must be separated by at least
      $T_\text{switch}^\text{RX}+T_\text{sweep}$.}
    \label{fig:rxclock}
\end{figure}

A global schedule is {\em feasible} if and only if it satisfies all feasibility
constraints at each transmitter, and all feasibility constraints at each
receiver.  (Recall that only the downlink is changed in fused LEO GNSS, so the
transmitter here is one SV beam, and the receiver is one user modem.)  Let
\emph{GNSS scheduler} refer to the subroutine of the global scheduler concerned
with fused LEO GNSS operations, and let the \emph{GNSS schedule} be its
product.  GNSS schedules are periodic.  One may visualize their period
$T_\text{period}$ as one revolution of the hand of a clock
(Figs.~\ref{fig:txclock} \& \ref{fig:rxclock}), with the schedule of ranging
bursts repeating after this interval.  The relative timing of events differs
between transmit and receive clocks due to time-of-flight effects.  Transmit
and receive constraints under which the GNSS scheduler operates are shown in
the two figures: bursts must not overlap at either the transmitter or the
receiver, but two bursts to the same cell (or two bursts from the same SV) may
be back-to-back.  However, bursts to different cells (or bursts from different
SVs) must be separated by a suitable interval so that the transmitter and
receiver can steer their antennas, and so that time skew across the cell
($T_\text{sweep}$) is taken into account.  If some users do not need PNT,
additional flexibility is possible; but this paper will make conservative
scheduling assumptions:
\begin{enumerate}[leftmargin=*]
    \item Every cell receives $n$ ranging bursts per $T_\text{period}$.
    \item Bursts and/or TX switching intervals from one beam-channel of an
        SV do not overlap in time.
    \item Bursts from one SV beam to different cells are separated by at
        least the TX switching interval $T_\text{switch}^\text{TX}$.
    \item Bursts and/or RX switching intervals on one channel do not overlap
        in time from any viewpoint in the target cell.
    \item Bursts to one channel in one cell from different SVs are separated by
        $T_\text{switch}^\text{RX}$ from any viewpoint in the target cell.
    \item TX switching events are separated by at least $T_\text{set-up}^\text{TX}$.
        \label{enum:txswitch}
    \item RX switching events are separated by at least $T_\text{set-up}^\text{RX}$.
        \label{enum:rxswitch}
    \item Bursts to neighboring cells on the same channel are non-overlapping in
        time.
        \label{enum:neighborrule}
\end{enumerate}

This paper assumes, based on the implementation of serial-to-parallel
converters using, e.g., paged register files, that constraints
\ref{enum:txswitch} and \ref{enum:rxswitch} do not apply in the case of {\em
switching back to the most recent coefficients}.  That is, the array may be
switched to new coefficients without erasing the old coefficients from its
memory.  Under this assumption, switching forward and then switching back
requires only the short interval $2T_\text{switch}$, rather than the longer
interval $2T_\text{switch} + T_\text{set-up}$.

A GNSS schedule consists of the following data:
\begin{enumerate}[leftmargin=*]
    \item $T_\text{period}$
    \item For each cell $k$, and for each ranging signal $s=1\ldots n$, a tuple (SV,
      channel, time modulo $T_\text{period}$, $T_\text{flight}$,
        $T_\text{sweep}$) indicating that a ranging burst will be sent by this
      SV on this channel towards this cell at such times.
\end{enumerate}
Timing information may be quantized to a resolution of \Tquantize\ without any
significant loss of scheduling flexibility.  Each SV must be provided with all
assignment tuples involving either itself or any of its primary cells.  The
time-of-flight data in secondary (cross-cell) assignment tuples allows each SV
to determine when it is safe to schedule data bursts.  Optionally, the timing
and time-of-flight parameters for primary ranging bursts may be left out of the
GNSS schedule, to be determined by local scheduling on each individual SV.

\subsubsection{Optimization}
The GNSS scheduler must be an efficient algorithm for finding feasible global
schedules.  Among feasible schedules, it will also optimize: the GNSS
scheduler will prefer assignments that minimize the impact on local data
scheduling, maximize the geometric diversity of ranging signals in each cell,
and minimize system power consumption.  The relative importance of these
objectives could depend on the prices paid by customers for broadband and PNT
service, and on the number of customers of each type in each cell.  Given the
complexity and non-convexity of this problem, approximations or heuristics will
likely be required.

\subsubsection{Complexity}
\begin{figure*}[htp]
  \centering
  \includegraphics[width=.8\textwidth]{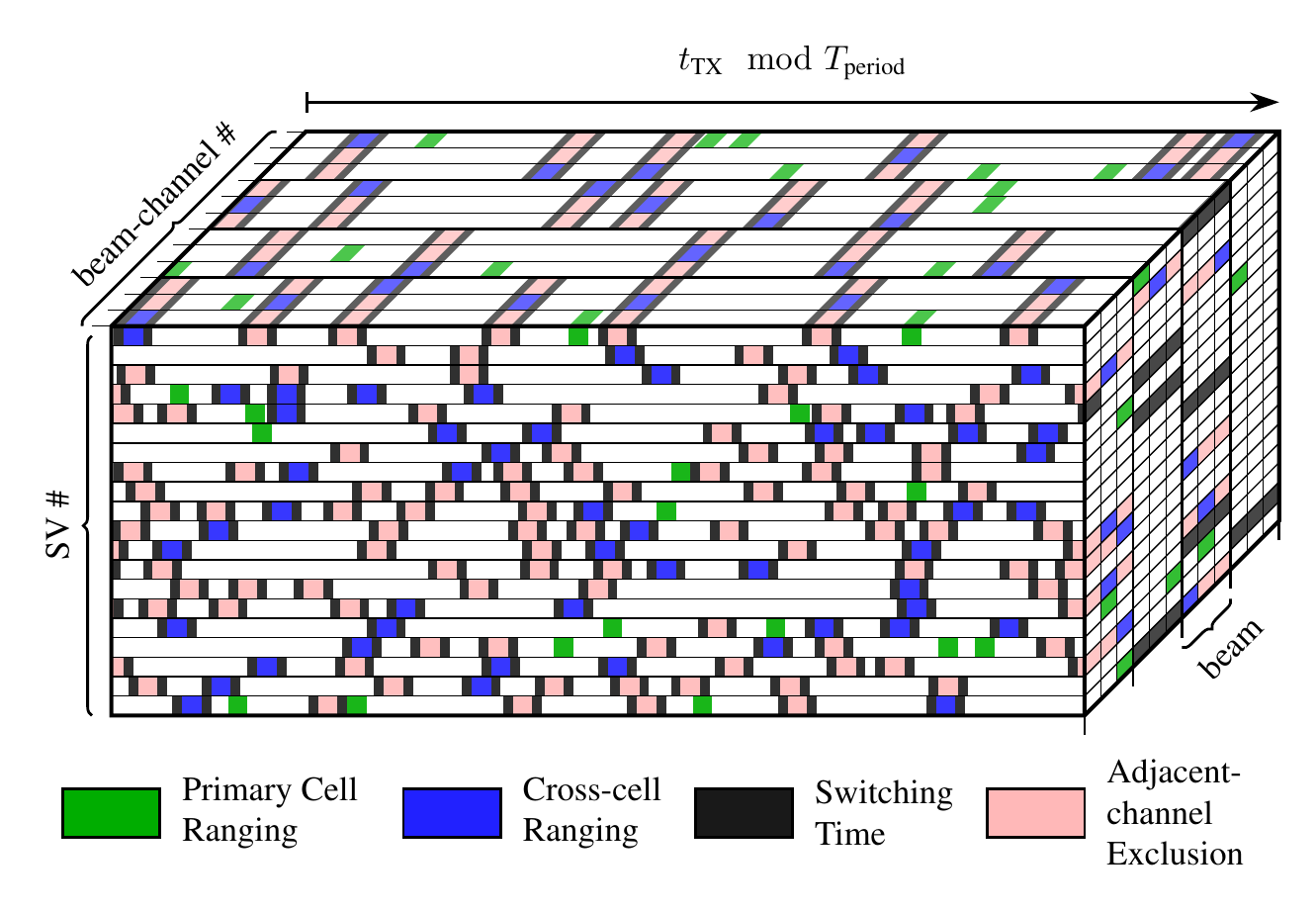}
  \caption{The TX cube.  The GNSS scheduler uses such a representation to
    track when each (beam, channel) pair of each SV has been reserved for
    ranging.  Burst durations and cube occupancy are greatly exaggerated for
    clarity of visualization.  Beam-channels belonging to a single beam are
    shown separated by thicker lines.}
  \label{fig:txcube}
\end{figure*}

\begin{figure*}[htp]
  \centering
  \includegraphics[width=.8\textwidth]{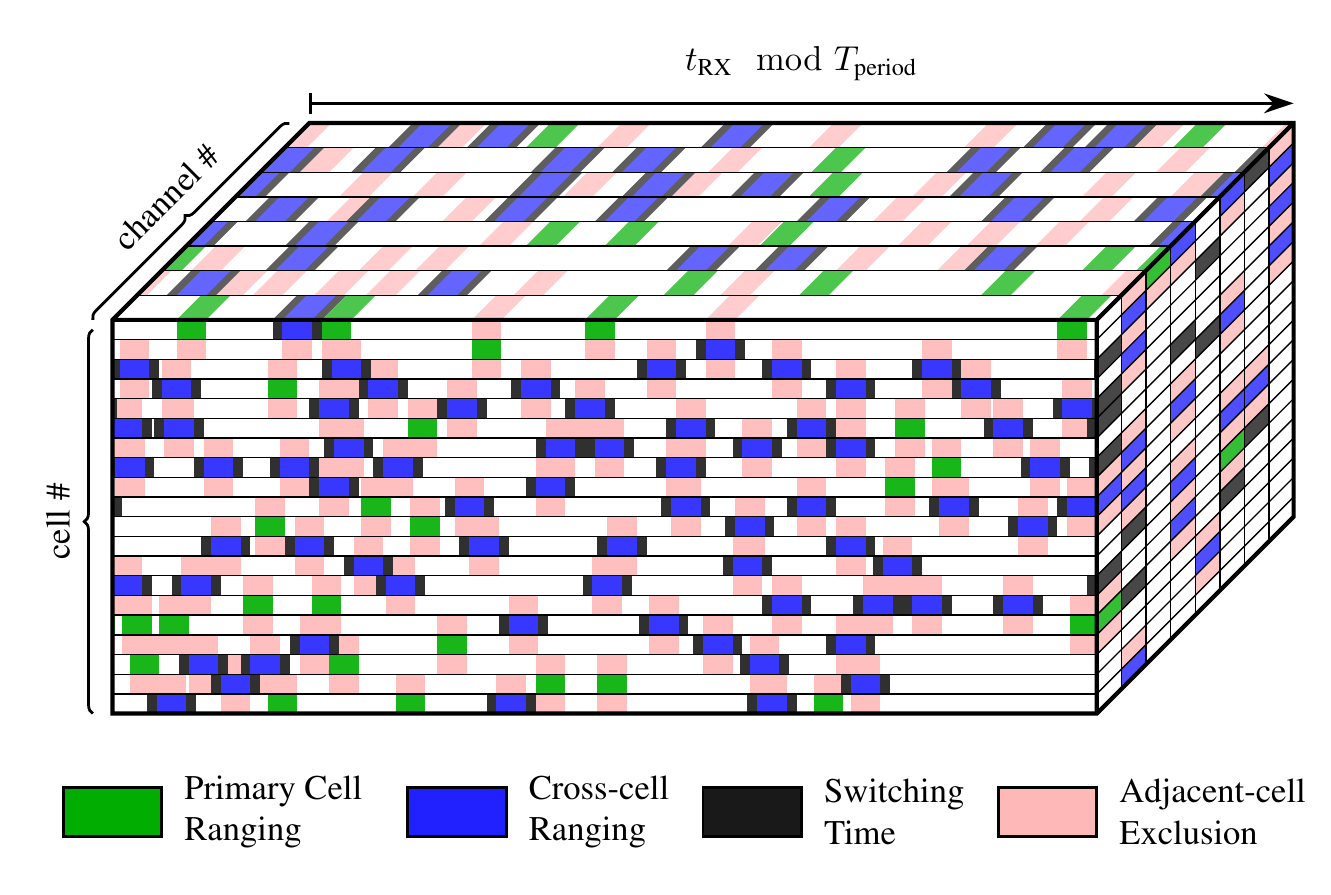}
  \caption{The RX cube.  The GNSS scheduler uses such a representation to
    track when each channel in each cell has been reserved for ranging.  The
    true RX cube has a more complicated adjacency relationship than shown
    here, because service cells form a 2-D hexagonal grid rather than a 1-D
    line.  Burst durations and cube occupancy are greatly exaggerated for
    clarity of visualization.}
  \label{fig:rxcube}
\end{figure*}

How complex is the global scheduling problem, and how over- or
under-constrained is the feasibility problem?  One way to grapple with this
question is to consider a simple greedy baseline GNSS scheduler that runs after
the global broadband scheduler has made its assignment of primary SV beams to
cells over the next planning window.  The GNSS scheduler allocates a ``transmit
cube'' array to represent the ranging-specific status of SV $i$, beam-channel
$bc$ at
time $t\mod T_\text{period}$ (Fig.~\ref{fig:txcube}).  Recalling that data
service is scheduled locally, and hence does not appear in the global schedule,
the enumerated status values stored in the TX cube are Idle, Burst, and Switch.
The GNSS scheduler also allocates a ``receive cube'' array to represent the
ranging-specific status of cell $k$, channel $j$ at time
$t\mod T_\text{period}$ (Fig.~\ref{fig:rxcube}).  The enumerated status values
stored in the RX cube are Idle, Burst, Switch, and Exclude.  During
construction of the TX and RX cubes, the GNSS scheduler consults an
``availability'' array, which indicates whether SV $i$ is available to provide
signals to cell $k$, or is excluded from doing so for any reason.

\begin{figure}[htp]
  \centering
  \includegraphics[width=\columnwidth]{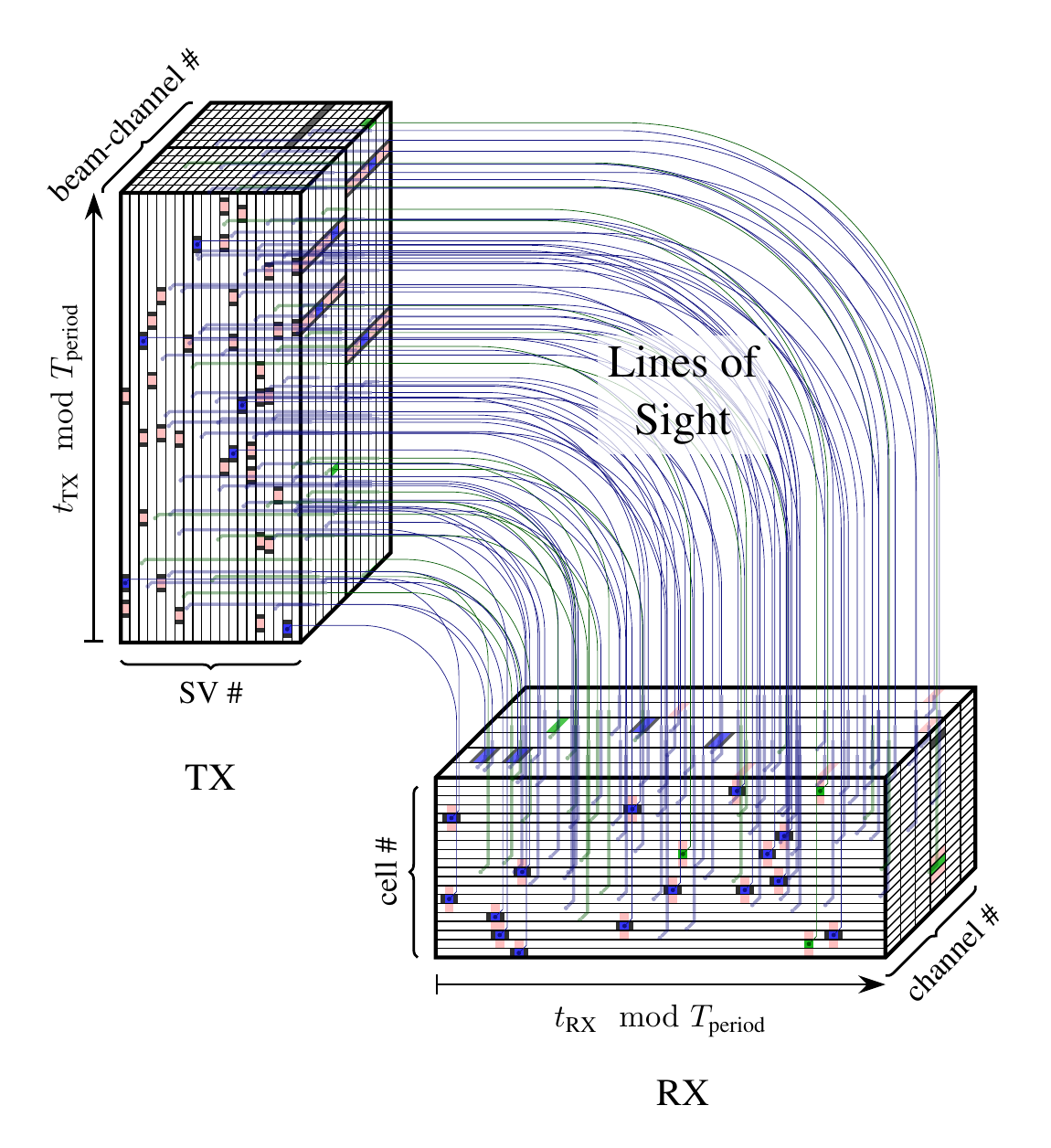}
  \caption{Notional schedule for ranging bursts on the coupled TX and RX cubes.
    A feasible schedule (\S\ref{ss:schedulefeasibility}) must satisfy four
    classes of requirements.  {\em Mutual consistency}: Each ranging burst must
    appear in both cubes with corresponding channel and time indices, respecting
    times-of-flight.  {\em Mutual visibility}: Each burst must arrive above the
    minimum elevation angle.  {\em Local consistency}: Each burst must
    satisfy the requirements particular to each cube.  {\em Global
    satisfaction}: An adequate number of bursts must reach each cell.  Each
    line-of-sight connects corresponding allocations in the TX and RX cubes, and
    is color-coded according to whether it is primary-cell (green) or cross-cell
    (blue).  Burst durations are greatly exaggerated for clarity of
    visualization.}
  \label{fig:txrxhypercube}
\end{figure}

The greedy baseline GNSS scheduler iterates over cells $k$.  For each cell, it
iterates over ranging signals $s=1\ldots n$.  For each signal, it iterates over
available SVs $i$.  To maximize geometric diversity for $n=5$, a ``goal
direction'' is defined for each $s$, and available SVs are iterated in
descending order of alignment of their line-of-sight vectors with the goal
direction.  The goal directions form a square pyramid of points in the sky
\cite{teng2016closed}: for $s=1$, the zenith; for $s=2$, local north; for
$s=3,4,5$, local east, south, and west, respectively.  The greedy scheduler
proceeds from SV to SV in this order, attempting to find idle space in the
corresponding planes of the TX and RX cubes to make an assignment.  When it
finds a suitable space, it adds the assignment tuple to the output GNSS
schedule, updates the TX and RX cubes, and proceeds to the next signal $s+1$.
If all available SVs are exhausted and $n$ signals have not been assigned for
cell $k$, then the greedy scheduler fails.  If all cells have been assigned $n$
signals each, then the greedy scheduler succeeds.
%

Is this procedure likely to succeed?  Does the answer depend upon whether cells
are iterated in geographic order, or randomly?  Or upon whether iteration is
instead first over signals and then over cells?  A key clue is that the TX and
RX cubes remain overwhelmingly Idle in any event, as will be shown.

\subsubsection{Schedule Sparsity}
\label{sec:schedule-sparsity}
If one pre-supposes that a feasible schedule exists, it consists of
$N_\text{cells}\cdot n$ assignment tuples, of which a fraction $(n-1)/n$ are
secondary, and the rest are primary.  The fraction of the TX cube that is
non-Idle is then equal to the sum duration of these assignments, divided by the
dimensions of the cube.  This fraction, which quantifies transmitter resources
devoted to the mega-constellation's secondary PNT mission and thus rendered
unavailable for its primary communications mission, will be called the {\em
  transmit reservation} $R_{\mathrm{TX}}$.  Similarly, the fraction of the RX
cube that is non-Idle will be called the {\em receive reservation}
$R_{\mathrm{RX}}$.

Each primary assignment reserves a single downlink SV beam-channel for an
interval $T_\text{burst}$.  Each secondary assignment reserves the entire SV
beam for $T_\text{burst}+2T_\text{switch}^\text{TX}$.  Summing up all the
primary and secondary assignments and dividing by the TX cube volume
$V_\text{TX}=N_\text{bc}\, N_\text{sats}\, T_\text{period}$, one obtains TX reservation
\begin{align}
  \label{eq:tx_res}
    R_{\mathrm{TX}} &\leq
    \frac{N_\text{cells}\left[T_\text{burst} + (n-1) (T_\text{burst} + 2 T_\text{switch}^\text{TX}) 
    \frac{N_\text{bc}}{N_\text{beams}}
    \right]}{N_\text{bc}\, N_\text{sats}\, T_\text{period}}
\end{align}
where the inequality allows for possible over-counting of switching times, and
non-uniform beam bandwidth is ignored.

Similarly, each primary or secondary assignment reserves one downlink channel
in a cell for an interval of up to $T_\text{burst}+T_\text{sweep}$, and each
secondary assignment additionally reserves the channel for
$2 T_\text{switch}^\text{RX}$.  Finally, constraint \ref{enum:neighborrule}
requires that each cell's $N_\text{adj}$ adjacent cells be reserved
for time $T_\text{burst}+T_\text{sweep}$ for each assignment.  Summing over
assignments and dividing by the RX cube volume
$V_\text{RX}=N_\text{cells}\,N_\text{channels}\, T_\text{period}$, one obtains
\begin{align}
  \label{eq:rx_res}
    R_{\mathrm{RX}} &\leq \frac{n(N_\text{adj}+1)(T_\text{burst}+T_\text{sweep})+2(n-1)T_\text{switch}^\text{RX}}{N_\text{channels}\, T_\text{period}} \\
    T_\text{sweep} & \leq \tfrac{D}{c} \cos(\phi_0),
\end{align}
where the first inequality allows for possible over-counting of switching times
and adjacent-cell exclusions, $\phi_0$ is the minimum elevation mask angle, $D$
is the cell diameter, and $c$ is the speed of light.

To obtain estimates for these reservations in practice, let
$N_\text{cells}=\SI{\Ncells}{}$, $N_\text{adj}=\Nadj$,
$N_\text{sats}=\SI{\Nsats}{}$, $n=\n$, $N_\text{channels}=\Nchannels$,
$N_\text{beams}=\Nbeams$, $T_\text{burst}=\Tburst$, $T_\text{switch}=\Tswitch$,
and $T_\text{period}=\Tperiod$.  For $T_\text{sweep}$, assume a minimum
elevation angle $\phi_0=\minElevation$ and a cell diameter $D=\D$.
(Other parameter values may be explored using the code in \ref{supp:costs}).
Then
\begin{align*}
    R_{\mathrm{TX}} \leq \TXreservationMax,\quad
    R_{\mathrm{RX}} \leq \RXreservationMax
\end{align*}

That the TX and RX cubes are overwhelmingly Idle is significant because the
scheduling problem within each cube is equivalent to the graph coloring problem
\cite{welsh1967upper} for which upper bounds exist on the number of required
colors (analogous to the number of time slices of the cube) for various
algorithms \cite{zufferey2008graph}.  So long as the cube reservations are
sparse, feasible schedules can be found with even very simple algorithms
\cite{welsh1967upper}.  One may thus presume that the feasibility problem is
greatly under-constrained, and the optimization problem will be of greater
interest than the feasibility problem in future work.

The global TX and RX reservations also equal the mean reservations for any
given SV beam-channel or cell, respectively.  Thus, the impact on the local scheduling
problem should be small as well.  This paper accordingly does not elaborate on
local scheduling.

One might question whether all or only a subset of SVs in a broadband LEO
system ought to be involved in fused LEO GNSS.  Indeed, the low reservation
numbers indicate that such sub-setting would be possible.  However, due to
switching times, it is less costly for an SV to provide ranging bursts to its
primary cells than otherwise.  It is preferable, then, to task every SV having
one or more primary cells to provide ranging service to those cells.  Load
shifting over regions with few subscribers could mean that some SVs are placed
into power-saving modes, while the remainder are tasked close to 100\%
capacity.  In this case, GNSS duties would also be concentrated as much as
possible to minimize the number of active SVs.


\subsection{Orbit and Clock Determination}
\label{ss:pod}
Both hosted payload and fused approaches to LEO GNSS require continual,
highly-accurate estimates of the SV orbital ephemerides and of the time offsets
of the space-borne clocks.  Such estimation should ideally take advantage of
constellation-to-ground ranging, intra-constellation ranging, and onboard GNSS
receivers. In the near term, however, broadband LEO providers should not be
expected to build out an observational network of ground stations extensive
enough (e.g., covering ocean regions), or intra-constellation ranging accurate
enough, to significantly improve on-orbit and clock determination based on
onboard GNSS receivers, which can be expected to constrain forecasting
uncertainties for clocks and orbits to \orbitclockrms{} RMS after one second,
as estimated in \cite{iannucci2020fused} based on Montenbruck et
al.~\cite{montenbruck2005reduced}.  Even better results are likely possible
using the more recent results of \cite{montenbruck2021performance,sun2017realtime}.

Thus, in the near term, LEO GNSS will operate in what Reid et al. refer to as a
multi-tier GNSS architecture \cite{Reid2016LeveragingCB}, with each SV carrying,
and dependent on, a GNSS receiver.  This architecture is intermediate in
assurance between traditional GNSS and fully autonomous LEO A-PNT: it is
reliant on GNSS being available in orbit, but provides highly jam-resistant
signals to users on the ground.

\section{Cost Model}
\label{ss:cost}

The opportunity cost of PNT provisioning for fused LEO GNSS is the value of the
best alternative use of the same resources.  In the context of this paper, the
opportunity cost of fused LEO GNSS is the value of {\em foregone broadband
service}.  To determine what quantity and value of broadband service is
foregone, one would need to know both the load on the system---since idle
resources cannot be wasted by putting them to profitable use---and the billing
structure for broadband data.  Neither is publicly known.  To address the
first issue, this paper will focus on {\em worst case} opportunity cost: lost
opportunities for profitable broadband service assuming the broadband LEO system
is {\em loaded at 100\% at all times}, i.e., that there is at least one
constellation resource that is 100\% utilized, the so-called bottleneck
resource.  To address the second issue, rather than giving costs in monetary
units, this paper will express costs as {\em reservations}: percentages of
potential bottleneck resources.

Lost opportunities for broadband data transmission may be interpreted
differently depending on the subscription model.  For a given number of users
with fixed load patterns, opportunity costs manifest as small penalties to
latency and throughput.  If the number of subscriptions is variable but
subscriber bandwidth and latency are held constant, opportunity costs
manifest as a reduction in the number of subscriptions that may be sold.  The
former model is more likely to apply for broadband LEO: guarantees in consumer
Internet service tend to be aspirational because quality of service can be more
nimbly modulated than number of subscribers.

This section will consider potentially-constraining resources one at a time,
supposing that each, in turn, is the bottleneck resource.  One presumes that a
prescient broadband LEO system operator would provision the constellation such
that each constraint is nearly binding in practice; otherwise, costs could have
been cut in some sub-system(s).  Discussion and interpretation of the results
follows in \S\ref{ss:discussion}.

\subsection{Scope}
The opportunity cost of a fused LEO GNSS service could be assesed for both
downlink and uplink, and for the SV modem, the user terminal (UT), and the
gateway terminal.  But not all link-endpoint combinations are independently
affected.  For example, LEO GNSS never prevents an SV from receiving uplinked
broadband data from UTs: SVs are assumed to be full duplex and they receive no
LEO-GNSS-specific signals from the UTs.  The only effect LEO GNSS has on uplink
reception involves command-and-control traffic from the gateway terminals, which
in (\S\ref{ss:uplinkcost}) is shown to be slight. 

From the perspective of a (presumably typical) half-duplex UT, time is the
limiting resource.  When the UT tunes to one channel, it cannot hear the
others; when it points towards one transmitter, it cannot decode another.  With
the introduction of a fused LEO GNSS service, non-participating UTs will notice
a slight reduction in available opportunities to receive downlink packets
(\S\ref{ss:downlinkcap}).  Participating UTs will additionally spend time
directing their array and/or tuning their receiver to capture scheduled ranging
bursts (\S\ref{ss:userterminal}).  However, if the UTs are not already 100\%
busy, fused LEO GNSS will have no effect on uplink transmission.

Thus, in what follows, SV downlink transmission resources will receive more
attention than SV uplink reception resources, and UT resources will only be
considered in terms of a lumped duty cycle.

\subsection{Downlink Capacity}
\label{ss:downlinkcap}
If an SV's local scheduling problem becomes over-constrained---that is, if no
more downlink data packets can be scheduled, but packets are available from the
gateway---then downlink capacity becomes a bottleneck resource.  In this case,
either data or ranging service is delayed, depending on what promises of
latency, throughput, or reliability are in effect.  (One presumes that to avoid
excessive buffering some form of active queue management is needed at the
gateway \cite{gettys2011bufferbloat}.)

At most, every channel of every beam of every satellite can be active, and at
most, every channel of every cell can be active.  The maximum downlink
throughput of the LEO broadband system is subject to both limits.

A broadband LEO downlink without fused LEO GNSS can thus deliver no more than
\begin{align}
    C^\text{DL}_\text{before} & = \frac{\min\{V_\text{TX}, V_\text{RX}\}}{T_\text{period}}
    \label{eq:cdlbefore}
\end{align}
channels' worth of data at once.  A ``channel's worth'' is one spatial degree of
freedom---one beam of a satellite or one cell---times one unit of bandwidth
allocation.  (A subtler calculation would involve multi-colorings of the hexagon
graph; \eqref{eq:cdlbefore} is merely an upper bound.)

Converting \eqref{eq:cdlbefore} into bits per second requires a model of the
link budget, signal-to-noise ratio, and forward error correction scheme.
\ref{supp:costs} develops a concrete estimate for this conversion factor
assuming the use of Turbo codes~\cite{berrou1993turbo}.  Under the modeling
assumptions further detailed in the Supplement, each channel's worth of
bandwidth (modeled as \W{}) supports \Rturbo{} of data.

Introducing fused LEO GNSS decreases each argument of the $\min$ function in
\eqref{eq:cdlbefore} in proportion to the corresponding reservation.  The
remaining downlink capacity is
\begin{align}
    C^\text{DL}_\text{after} & = \frac{\min\{V_\text{TX}(1-R_{\mathrm{TX}}),
    V_\text{RX}(1-R_{\mathrm{RX}})\}}{T_\text{period}}
\end{align}
for a relative change of
\begin{align}
    \bar{R}_\text{DL} & = \frac{C^\text{DL}_\text{before} - C^\text{DL}_\text{after}}{C^\text{DL}_\text{before}} = \downlinkReservationMax
\end{align}
or an absolute change of \broadbandEquivalentService{} per cell.

\subsection{Scheduling Complexity}
To obtain a concrete bound on the computational complexity of global
scheduling, consider a greedy randomized scheduler.  The expected number of
attempts needed to pick an available tuple (SV, beam, channel, time) by
rejection sampling scales as $1/p$ with the probability $p$ of success, or the
probability that a randomly-chosen tuple of parameters points to unallocated
space within both the TX and RX cubes.  As more allocations are added to the
schedule, the empty space in the cubes decreases, and the chance that the next
randomly-sampled tuple of parameters will be feasible falls.  Let
$p_\text{min}$ be the probability of success for the last tuple to be
scheduled.  Then the expected complexity of building a schedule with
$n\,N_\text{cells}$ total allocations is no greater than
$n\,N_\text{cells}/p_\text{min}$.  By the union bound,
\begin{align}
    p_\text{min} \geq 1 - \mathbb{P}\{\text{TX cube collision}\} - \mathbb{P}\{\text{RX cube collision}\}.
\end{align}
These two quantities differ from the transmit and receive reservations by a
factor less than $2$ due to the need to avoid overlap between allocations of
finite extent.  One may therefore bound the expected time complexity of finding
a feasible schedule by
\begin{align}
    \frac{n N_\text{cells}}{1-2R_{\mathrm{TX}}-2R_{\mathrm{RX}}} =
    \expectedComplexityMax.
\end{align}
Each step requires sampling a random tuple, testing for collisions in each
cube, and potentially writing a new allocation into each cube.  These millions
of steps can be completed in less than a second on a modern processor.

\subsection{Phased Array Set-Up}
\label{ss:setup}
Another constellation resource consumed by fused LEO GNSS is serial bandwidth
for phased array set-up~(\S\ref{ss:beamforming}).  This section first estimates
a baseline for the SV's capacity to steer a beam, expressed through the
parameter $T_\text{set-up}$, from considerations of link budget and end-to-end
latency.  It then compares this quantity to the fused LEO GNSS schedule to
compute the consumed portion $\bar{R}_{\mathrm{SU}}$ of constellation steering
resources.

\subsubsection{$T_\text{set-up}$}
Broadband LEO SVs steer their arrays both to compensate for continuous orbital
motion and to switch between service cells.  Orbital motion changes the SV-UT
line of sight by up to $\omega=\broadbandSteeringRate$.  With a phased array,
steering occurs in discrete steps.  Suppose these occur at intervals
$T_\text{steer}$.  If steering events are infrequent, the steering error cannot
be kept small, and antenna gain along the line-of-sight fluctuates downward.
Thus, the communications link budget imposes an upper bound on $T_\text{steer}$.

Consider a \beamFWHM{}-wide beam (\ref{supp:sops}, Table I), measured in terms
of full width at half maximum (FWHM).  The decrease in antenna gain ({\em
pointing loss}) at an angle $\Delta\theta$ off-axis is
$L_\text{pointing}=\SI{12}{\dB}(\Delta\theta/\text{FWHM})^2$ under the Gaussian
approximation.  Given a loss budget $L_\text{pointing}^\text{max}$, the interval
between steering events must not exceed
\begin{align}
    T_\text{steer}^\text{max} & \leq \frac{2\Delta\theta}{\omega} = \frac{\text{FWHM}}{\omega}
    \sqrt{\frac{L_\text{pointing}^\text{max}}{\SI{3}{\dB}}}
\end{align}
If $L_\text{pointing}^\text{mean}$ is given rather than
$L_\text{pointing}^\text{max}$, the 3 in the denominator is replaced by 1.
An exact relationship between $\Delta\theta$ and $L_\text{pointing}$ is
discussed in \ref{supp:pointing}, following \cite{buchsbaum1986pointing}.
$T_\text{steer}$ can be further cast in terms of lost throughput: for instance,
to limit throughput losses due to misalignment to \pointingThroughputPenaltyA{},
$T_\text{steer}$ must be less than \pointingTsteermaxA{}.

The second cause of steering events is switching service cells.  If an SV serves
broadband to one cell per beam, there may be minutes between these switching
events.  However, an SV beam serving multiple cells must steer more frequently.
To reach a target latency---for Starlink, \SI{20}{ms}\cite{MuskTweetLatency}---a
user can never be neglected for a period approaching the end-to-end latency.
The downlink array must be steered more than once per \SI{20}{ms} for each cell
it serves with broadband.  This is expected to be case early in the deployment
of any mega-LEO constellation, when few SVs are available.  Suppose, then, that
downlink arrays are designed for a small multiple of 50 steering events per
second.  This is vastly more than $1/T_\text{steer}^\text{max}$, but only
sufficient for a small number of service cells per beam.  To pick a concrete
number, suppose each array can accept a new set of coefficients each
$T_\text{set-up}=\Tsetup$.

\subsubsection{$\bar{R}_{\mathrm{SU}}$}
Fused LEO PNT requires each array to be steered frequently to provide secondary
ranging bursts.  If the GNSS schedule is designed jointly with the
broadband schedule, so that a beam's broadband service assignments coincide with
its ranging service assignments, then the need for additional steering events
may be reduced.  Otherwise, this limits opportunities for assigning many cells'
broadband service to the same SV.

Recall (\S\ref{ss:schedulefeasibility}) the assumption that ``switching back''
has no set-up cost: during switching, the banks of memory
holding old and new coefficients are swapped without erasing, so that switching
back does not require re-loading coefficients.  Otherwise, the steering
bandwidth utilization must be doubled.

Steering bandwidth is the least abundant resource considered so far.  The GNSS
schedule includes a total of $(n-1)\,N_\text{cells}$ secondary assignments,
each requiring one SV beam's phasing array to be busy loading coefficients for
an interval $T_\text{set-up}$ once per $T_\text{period}$.  Considered as a
fraction of total steering set-up capacity in the same way as the TX
reservation, the mean set-up reservation $\bar{R}_{\mathrm{SU}}$ is given by
\begin{align}
  \label{eq:su_res}
  \bar{R}_{\mathrm{SU}} = \frac{(n-1)\,N_\text{cells}\,T_\text{set-up}}{N_\text{beams}\, N_\text{sats}\, T_\text{period}}  = \setupReservationMax
\end{align}

\subsection{Power}
\label{sec:power-management-cost}
Available electrical power becomes a bottleneck resource when the constellation
operates in a scarce energy regime (\S\ref{sec:power-management}).  Opportunity
cost of fused LEO GNSS attains a maximum in this regime: a Joule expended on a
ranging burst is a Joule that could have been spent on broadband service.
Moreover, if the SV's battery is sized so that peak power loads are amortized
across the orbit, then the cumulative energy expended per orbit drives the
calculation of opportunity cost, in which case it makes no difference at what
point in the orbit the ranging burst is transmitted, whether over a sparsely-
or densely-populated area: the opportunity cost per burst remains the same
(maximum) value for every burst.

As discussed in \S\ref{sec:power-management}, SVs will be commanded to enter a
``deep sleep'' state during some portion of their orbit, and otherwise may be
expected to operate near capacity.  The burden of fused LEO GNSS in any given
region is distributed among the active SVs.  Let $\bar{E}_\text{burst}$ be the
average energy expended by an SV to transmit a data or ranging burst via a
single channel of a single beam,
let $\bar{E}_\text{orbit}$ be the average energy expended by an SV's
transmitters per orbit, and let $T_\text{orbit}$ be the average orbital period.
The number of ranging bursts per orbit per satellite is 
$n\,N_\text{cells}\,T_\text{orbit}/T_\text{period}/N_\text{sats}$.  Then the
mean energy reservation $\bar{R}_{\mathrm{E}}$, or the mean energy allocated per
orbit for fused LEO GNSS as a fraction of total-constellation mean energy
expended for downlink, can be approximated as
\begin{align}
    \bar{R}_{\mathrm{E}}
    & = \frac{n\, N_\text{cells}\,T_\text{orbit}\,\bar{E}_\text{burst}}{T_\text{period}\,N_\text{sats}\,\bar{E}_\text{orbit}}
\end{align}
Without detailed knowledge of the broadband LEO system, is not possible to
estimate $\bar{E}_\text{burst}/\bar{E}_\text{orbit}$, the ratio of the modem's
energy consumption to transmit a single burst on a single channel of a single
beam to the total energy expended on downlink over an orbit.  A more accessible
parameter is the transmitter peak-to-average power ratio (PAR) for an SV: the
ratio of the power consumed when the downlink is operating at full capacity to
the power consumed by the downlink on average.  At maximum output, the
transmitters use
$P_\text{max}^\text{TX} =
\frac{N_\text{bc}\,\bar{E}_\text{burst}}{T_\text{burst}}$, whereas on average,
they use
$P_\text{mean}^\text{TX} = \frac{\bar{E}_\text{orbit}}{T_\text{orbit}}$.  This
implies
\begin{align}
    \text{PAR} & = \frac
        {P_\text{max}^\text{TX}}
        {P_\text{mean}^\text{TX}}
        = \frac
        {N_\text{bc}\,\bar{E}_\text{burst}\,T_\text{orbit}}
        {T_\text{burst}\,\bar{E}_\text{orbit}} \\
    \bar{R}_{\mathrm{E}} & = \frac
        {n\, N_\text{cells}\,T_\text{burst}\,\text{PAR}}
        {T_\text{period}\, N_\text{sats}\,N_\text{bc}} \label{eq:e_res}
\end{align}
Suppose SVs spend equal power on each visited subscriber so that, to a first
approximation, power is consumed in proportion to the number of
subscribers-in-view.  Then PAR can be predicted from the geographic
distribution of LEO broadband subscribers.  (This assumes that the average
power allocated to fused LEO GNSS is a small fraction of each SV's total
average power load.) The distribution of broadband subscribers is
thus a key input in estimating the power opportunity cost of fused LEO GNSS.

\subsubsection{Estimating the Geographic Subscriber Distribution}
In urban centers, terrestrial broadband Internet service enjoys better scaling
than broadband LEO service: terrestrial capacity can be expanded locally,
whereas LEO capacity can only be expanded globally.  For this reason, the
population distribution of the world is not a good estimate of the distribution
of potential subscribers.  Suppose the set of potential customers is primarily
rural and lacks access to terrestrial broadband alternatives.  How might the
distribution of such be estimated?

Let $\rho_\text{max}$ be the maximum population density that can be
competitively served by broadband LEO.  For instance, suppose that in regions
exceeding this density terrestrial broadband is available and inexpensive, and
satellite downlink capacity is saturated due to regulatory flux limits
(\S\ref{ss:flux}).  Let $\rho_\text{max}$ be estimated by
$\hat{\rho}_\text{max} = \gamma\cdot \hat{\rho}_\text{max}^\text{USA}$, where
$\hat{\rho}_\text{max}^\text{USA}$ is the density threshold in the United
States, and $\gamma$ accounts for international differences in
broadband usage per (potential) customer.

In \ref{supp:population}, $\hat{\rho}_\text{max}^\text{USA}$ is determined from
the estimated number of U.S. citizens without options for purchasing
terrestrial broadband Internet (42 million circa
2020)~\cite{broadbandNow2020unconnected}.  The Supplement supposes these
individuals to be precisely those residing in the lowest-density regions.  From
the Gridded Population of the World, Version 4
(GPWv4)~\cite{CIESIN_GPWv4_2018}, \ref{supp:population} extracts, for each cell
within the land area of the United States,
both the population density and the integrated population count.  It finds the
cumulative distribution of individuals by the density of their regions,
accumulating population counts in order from least dense to most dense.  This
cumulative distribution first exceeds the target threshold of 42 million people
at a density of \pmaxDeveloped{}.

\ref{supp:population} estimates $\gamma$ as the ratio of active mobile broadband
subscriptions per 100 inhabitants in the developed world to that in the entire
world.  That is, holding all else equal, if mobile broadband subscriptions are
more numerous on a per-inhabitant basis in some region, then the broadband LEO
system will tend to saturate (that is, reach its maximum sustainable
\si{subscribers\per\km^2}) at fewer inhabitants per \si{\km^2} there.  This
ratio is approximately 122/83 (circa 2019)~\cite{itu2019measuringFacts}, giving
a global average value of $\hat{\rho}_\text{max} = \pmax$.  This holds under the
simplistic assumption that mobile broadband penetration can be taken as a proxy
for LEO broadband demand.  The reality is more complicated, since high-rate
mobile broadband may be a substitute for broadband LEO; but a more precise value
would be unlikely to significantly change the result.


To account for limited system capacity in populous regions---which may
nevertheless contain a substantial number of
subscribers---\ref{supp:population} limits all regions to a maximum density
of $\hat{\rho}_\text{max}$.  A grid $p'_{ij}$ of potential-subscriber counts is
computed from the GPWv4 population density grid, and then summed over the
circular neighborhood (correcting for spherical geometry) of every potential
satellite location.  This yields a visible-subscriber count.  Finally, this
count array is sampled over various orbital parameters to obtain a distribution,
from which peak and average values may be computed.

For a LEO constellation at an altitude of \SI{550}{\km} and an orbital
inclination of \SI{53}{\degree} (Starlink's proposed B
sub-constellation\cite{starlink2019parameters}), the peak-to-average population
density ratio resulting from such calculations is \PARmin\
(\ref{supp:population}), which may be taken as a proxy for PAR.  Filling in the
other quantities involved in $\bar{R}_{\mathrm{E}}$ with the values given in
\S\ref{sec:schedule-sparsity}, one finds that the mean energy reservation is
$\bar{R}_{\mathrm{E}} = \powerReservationMax$.

\subsection{Command-and-Control Bandwidth}
\label{ss:uplinkcost}
Fused LEO GNSS will require only a negligible amount of command-and-control
bandwidth.  This paper assumes that each SV independently performs
precision orbit determination using an on-board GNSS receiver.  This requires
streaming traditional GNSS satellites' precise orbit and clock models from the
gateway.  Such a stream could be derived from, for instance, the IGS Real Time
Service\cite{agrotis2017igsrts,igs2020annual,sturze2020igsrts}, which consumes
little bandwidth: \SIrange{400}{800}{\bit\per\sec} for precise orbit and clock
corrections including every constellation, \SI{3.4}{\bit\per\sec} for global
ionospheric model coefficients, and optionally \SI{17}{\kilo\bit\per\sec} of
broadcast ephemeris data for navigation bit wipe-off.

The LEO GNSS schedule must also be distributed, sending each SV its assignments
and those secondary assignments which affect its primary cells
(\S\ref{ss:schedulefeasibility}).  Overall, each primary assignment in the GNSS
schedule will be sent to exactly one SV, and each secondary assignment will be
sent to exactly two: the transmitter, and the SV that is primary for the target
cell.  The total bandwidth used distributing assignments is therefore
$(2n-1)\,N_\text{cells}$ times the size in bits of one assignment.  Examining
the range and quantization of each parameter in an assignment tuple, one finds
that the tuple may be encoded in \assignmentSizeBits\ bits
(\ref{supp:costs}).  The uplink bandwidth used by fused LEO GNSS is
dominated by the cost of uplinking assignments.  Summed over the entire
constellation, this assignment uplink cost may be computed as
\begin{align}
  \label{eq:assignmentUplinkCost}
   C_{\mathrm{AU}}  = (2n-1)\,N_\text{cells}\cdot \SI{\assignmentSizeBits}{b}
    \lesssim\assignmentUplinkCost
\end{align}
The data portion of each primary ranging burst will include a copy of the
secondary assignments.  Users also need the line-of-sight direction to each
secondary SV for steering.  It is sufficient to provide orbits accurate to
$\sim\SI{10}{km}$ (\ref{supp:pointing}) for this purpose: once a
secondary burst is decoded, the user will have high-accuracy ephemeris for that
SV.

These assignments and ephemerides need to be updated on the SV no more than a
few times per minute.  On a per-SV basis, the total uplink cost is approximately
\uplinkRatePerSV{} (\ref{supp:costs}).

\subsection{User Terminal Duty-Cycle}
\label{ss:userterminal}

For reasons of cost, full-duplex user terminals seem a remote possibility.
Consider, then, a half-duplex UT that divides its time between uplink, downlink,
switching between the two, and sitting idle.  Let the duty cycles of these
activities be expressed as
\begin{align}
    d^\text{UT}_\text{UL} + d^\text{UT}_\text{DL} + d^\text{UT}_\text{Switch} + d^\text{UT}_\text{Idle} & = 100\%
    \label{eq:rxtimebudget}
\end{align}

From the perspective of an individual UT, there are three cases to consider with
regards to fused LEO GNSS service: service may be absent, unneeded, or in use.
If service is absent, the UT may spend up to 100\% of its time in either of the
uplink or downlink states, achieving maximum data throughput.

If service is present, then whether or not it is needed by this UT, the
downlink channel is partly occupied, and hence unavailable for other purposes.
The downlink duty cycle for an individual UT, $d^\text{UT}_\text{DL}$, is then
bounded above by $1-R_\text{RX}=\RXremaining{}$, and the average downlink duty
cycle for all UTs, $\bar{d}\,^\text{UT}_\text{DL}$, is bounded above by
$1-\bar{R}_\text{DL}=\downlinkRemaining{}$.  Uplink, switching, and idle duty
cycles are not directly impacted unless the UT is listening for fused LEO GNSS.

If an individual UT is using fused LEO GNSS, then a cost $d_\text{PNT}$ is
subtracted from the right-hand side of \eqref{eq:rxtimebudget}, with
\begin{align}
    d_\text{PNT} & \leq \frac{n T_\text{burst} + 2 (n-1) T_\text{switch}^\text{RX}}{T_\text{period}} = \dpnt{}
\end{align}
This is a less stringent limitation than the former bounds on downlink duty
cycle, but it directly bounds $d^\text{UT}_\text{UL} \leq \dUTULMax{}$.

\subsection{Discussion}
\label{ss:discussion}
Regarding the mean downlink reservation $\bar{R}_{\mathrm{DL}}$: If the
broadband data load on an SV keeps its transmitters busy less than
\downlinkRemaining\ of the time, then $\bar{R}_{\mathrm{DL}}$ represents
essentially zero opportunity cost.  Recall that ranging bursts do not involve
gateway or inter-satellite link traffic.  It may be that gateway
retransmissions or temporary link interruptions cause the SV's packet buffers
to be empty more than \downlinkReservationRound\ of the time, even under heavy
offered broadband load.  If buffers are kept deliberately small for low
latency\cite{MuskTweetLatency,gettys2011bufferbloat}, link utilization might be
bounded away from 100\% due to the behavior of TCP.  Like water poured into the
interstitial spaces in a jar of sand, the channel reservation for fused
LEO GNSS ranging bursts may not displace any data traffic at all.  In any event,
a global schedule providing $n=\n$ ranging signals per second to every cell
between $\pm\serviceLatMax$ latitude would tie up no more than
\downlinkReservationRound\ of system downlink capacity.  This allocation is
comparable to adding one user consuming \broadbandEquivalentService\ of
broadband service to each cell (\ref{supp:costs}).

The mean energy reservation $\bar{R}_{\mathrm{E}}$, at \powerReservationMax{}
assumes the SV is capable of transmitting on all $N_\text{bc}=\Nbc$
beam-channels at once.  If this is not true, it would imply a larger value of
$\bar{E}_\text{burst}/\bar{E}_\text{orbit}$ and hence $\bar{R}_{\mathrm{E}}$.
The code in \ref{supp:costs} permits exploration of alternative scenarios.

The mean set-up reservation $\bar{R}_{\mathrm{SU}}=\setupReservationMax{}$
appears large, but its impact is subtle.  Broadband LEO service does not fail
catastrophically when beam steering is overtaxed.  Instead, the link budget
gradually degrades as beams spend more time out of perfect alignment.
Spill-over into adjacent cells also increases.  These potential problems are
reduced if cross-cell ranging bursts are timed to coincide with steering
updates, though this is not a complete solution.  There is not enough time to
load fresh coefficients for cell A while ranging is served to cell B if
$T_\text{set-up} > T_\text{burst}+2 T_\text{switch}^\text{TX}$.  As discussed
previously, if the array hardware permits recalling one or more sets of
coefficients rather than reloading them from the CPU, switching costs can be
substantially mitigated.  In this case, the strategy would be to begin loading
coefficients further in advance of the cross-cell burst.

In any event, as indicated in \S\ref{ss:setup}, the bulk of switching events for
broadband service are for time multiplexing, not for maintaining the link
budget.  Foregone opportunities to load new coefficients due to fused LEO
resource utilization lead to minuscule pointing-related throughput losses of
\pointingThroughputPenaltyB{}.  The true impact of $\bar{R}_{\mathrm{SU}}$ is
that an SV tasked with fused LEO GNSS cannot afford to time-multiplex between so
many cells as an SV free of fused LEO assignments.  Fortunately, the size of
this effect will diminish as mega-LEO constellations approach full utilization,
since time multiplexing between cells is not helpful if a single cell consumes
100\% of a beam's throughput.

%
%
%
%
%
%
%
%
%
%
%
%
%
%
%

The foregoing mean reservations have all been
calculated assuming a global ($\pm\serviceLatMax$ latitude) fused LEO GNSS
service, without regard to the distribution of users.  An alternative model
would provide service only to areas where subscribers are located.  For even
greater flexibility in matching supply with demand, such targeted service could
be paired with time-varying subscription rates.  Under this model, inflexible
customers demanding continuous high-accuracy LEO GNSS could obtain it, but at a
significant cost during periods and within regions of peak broadband demand.
Conversely, subscribers willing to accept opportunistic LEO GNSS service could
obtain it cheaply when and where its provision presents a near-zero marginal
cost to the broadband LEO system.

\section{Conclusion}
This paper presented a concept of operations for fused LEO GNSS, enabling the
exploitation of powerful new broadband LEO constellations for global
positioning, navigation, and timing (PNT).  It laid out a summary and analysis
of what is publicly known and what may reasonably be inferred about broadband
LEO systems, insofar as this information is needed to explore dual-purposing
these systems for PNT.  Finally, it analyzed the opportunity cost to
constellation providers for re-allocating resources to provide a fused LEO GNSS
service.  For a constellation such as SpaceX's Starlink, to provide continuous
service to \serviceRegionPopFraction{} of the world's population would require
reserving at most \downlinkReservationRound{} of system downlink capacity,
\powerReservationMax{} of system energy capacity, and \uplinkRatePerSV{} per SV
of command-and-control bandwidth.  This provisioning scenario reserves
\setupReservationMax{} of the constellation's capacity for beam-steering,
limiting the number of cells served by each SV beam.

\section*{Acknowledgments}
The authors wish to thank Lakshay Narula and Daniel LaChapelle for valuable
assistance.  Research was sponsored by the Army Research Office under
Cooperative Agreement W911NF-19-2-0333. The views and conclusions contained in
this document are those of the authors and should not be interpreted as
representing the official policies, either expressed or implied, of the Army
Research Office or the U.S. Government. The U.S. Government is authorized to
reproduce and distribute reprints for Government purposes notwithstanding any
copyright notation herein.


\makeatletter
\newcommand\supplement[3]{
\def\@currentlabel{Supplement~#1}%
\vskip.2pt{\leavevmode%
\nobreak\hskip-1em{\hbox to1.75em{\numberline{#1}\hfil}#2}\nobreak\dotfill\nobreak{\tt #3}\par}
}
\makeatother
\par
\section*{Index of Supplementary Material}
\supplement{A}{Costs, Geometry, Link Budget}{supplement\_a.py}\label{supp:costs}
\supplement{B}{PAR Estimate}{supplement\_b.py}\label{supp:population}
\supplement{C}{Antenna Pointing}{supplement\_c.pdf}\label{supp:pointing}
\supplement{D}{Signals-of-Opportunity}{supplement\_d.\{pdf,py\}}\label{supp:sops}

\bibliographystyle{ieeetran} 
\afterpage{\afterpage{\balance}}
\bibliography{pangea}

\begin{thebibliography}{10}
\providecommand{\url}[1]{#1}
\csname url@samestyle\endcsname
\providecommand{\newblock}{\relax}
\providecommand{\bibinfo}[2]{#2}
\providecommand{\BIBentrySTDinterwordspacing}{\spaceskip=0pt\relax}
\providecommand{\BIBentryALTinterwordstretchfactor}{4}
\providecommand{\BIBentryALTinterwordspacing}{\spaceskip=\fontdimen2\font plus
\BIBentryALTinterwordstretchfactor\fontdimen3\font minus
  \fontdimen4\font\relax}
\providecommand{\BIBforeignlanguage}[2]{{%
\expandafter\ifx\csname l@#1\endcsname\relax
\typeout{** WARNING: IEEEtran.bst: No hyphenation pattern has been}%
\typeout{** loaded for the language `#1'. Using the pattern for}%
\typeout{** the default language instead.}%
\else
\language=\csname l@#1\endcsname
\fi
#2}}
\providecommand{\BIBdecl}{\relax}
\BIBdecl

\bibitem{GPWv4}
\BIBentryALTinterwordspacing
{Center for International Earth Science Information Network - Columbia
  University}, ``{Gridded Population of the World, Version 4 (GPWv4):
  Population Count, Revision 11},'' Palisades, NY, 2018. [Online]. Available:
  \url{https://doi.org/10.7927/H4JW8BX5}
\BIBentrySTDinterwordspacing

\bibitem{morton2020position}
Y.~J. Morton, F.~van Diggelen, J.~J. Spilker~Jr, B.~W. Parkinson, S.~Lo, and
  G.~Gao, \emph{Position, Navigation, and Timing Technologies in the 21st
  Century, Volumes 1 and 2: Integrated Satellite Navigation, Sensor Systems,
  and Civil Applications, Set}.\hskip 1em plus 0.5em minus 0.4em\relax John
  Wiley \& Sons, 2020.

\bibitem{j_spilker96_atp}
J.~J. Spilker, Jr., \emph{Global Positioning System: Theory and
  Applications}.\hskip 1em plus 0.5em minus 0.4em\relax Washington, D.C.:
  American Institute of Aeronautics and Astronautics, 1996, ch. 3: {GPS} Signal
  Structure and Theoretical Performance, pp. 57--119.

\bibitem{j_spilker96_int}
------, \emph{Global Positioning System: Theory and Applications}.\hskip 1em
  plus 0.5em minus 0.4em\relax Washington, D.C.: American Institute of
  Aeronautics and Astronautics, 1996, ch. 20: Interference Effects and
  Mitigation Techniques, pp. 717--771.

\bibitem{teunissen2017springer}
P.~Teunissen and O.~Montenbruck, Eds., \emph{Springer handbook of global
  navigation satellite systems}.\hskip 1em plus 0.5em minus 0.4em\relax
  Springer, 2017.

\bibitem{dolman2012new}
E.~C. Dolman, ``New frontiers, old realities,'' \emph{Strategic Studies
  Quarterly}, vol.~6, no.~1, pp. 78--96, 2012.

\bibitem{washTimesRaymond2021}
\BIBentryALTinterwordspacing
B.~Gertz, ``{Air Force Gen. John W. Raymond: Chinese lasers, jammers threaten
  GPS satellites},'' June 2021. [Online]. Available:
  \url{https://www.washingtontimes.com/news/2021/may/10/air-force-gen-john-w-raymond-chinese-lasers-jammer/}
\BIBentrySTDinterwordspacing

\bibitem{humphreysGNSShandbook}
T.~E. Humphreys, ``Interference,'' in \emph{Springer Handbook of Global
  Navigation Satellite Systems}.\hskip 1em plus 0.5em minus 0.4em\relax
  Springer International Publishing, 2017, pp. 469--503.

\bibitem{psiakiNewBlueBookspoofing}
M.~L. Psiaki and T.~E. Humphreys, \emph{Position, Navigation, and Timing
  Technologies in the 21st Century: Integrated Satellite Navigation, Sensor
  Systems, and Civil Applications}.\hskip 1em plus 0.5em minus 0.4em\relax
  Wiley-IEEE, 2020, vol.~1, ch. {C}ivilian {GNSS} {S}poofing, {D}etection, and
  {R}ecovery, pp. 655--680.

\bibitem{luba2005gps}
O.~Luba, L.~Boyd, A.~Gower, and J.~Crum, ``{GPS} {III} system operations
  concepts,'' \emph{IEEE Aerospace and Electronic Systems Magazine}, vol.~20,
  no.~1, pp. 10--18, 2005.

\bibitem{t_humphreys08_cfi}
T.~E. Humphreys, L.~Young, and T.~Pany, ``Considerations for future {IGS}
  receivers,'' in \emph{Position Paper of the 2008 {IGS} Workshop}, 2008.

\bibitem{iannucci2020fused}
P.~A. Iannucci and T.~E. Humphreys, ``Economical fused {LEO} {GNSS},'' in
  \emph{Proceedings of the IEEE/ION PLANSx Meeting}, 2020.

\bibitem{levanon1998quick}
N.~Levanon, ``Quick position determination using 1 or 2 {LEO} satellites,''
  \emph{IEEE Transactions on Aerospace and Electronic Systems}, vol.~34, no.~3,
  pp. 736--754, 1998.

\bibitem{rabinowitz2000some}
M.~Rabinowitz, B.~Parksinson, and J.~Spilker, ``Some capabilities of a joint
  gps-leo navigation system,'' in \emph{Proceedings of the 13th International
  Technical Meeting of the Satellite Division of The Institute of Navigation
  (ION GPS 2000)}, 2000, pp. 255--265.

\bibitem{lawrence2016test}
D.~Lawrence, H.~S. Cobb, G.~Gutt, F.~Tremblay, P.~Laplante, and M.~O’Connor,
  ``Test results from a leo-satellite-based assured time and location
  solution,'' in \emph{Proceedings of the 2016 International Technical Meeting
  of The Institute of Navigation}, 2016, pp. 125--129.

\bibitem{khalife2019receiver}
J.~J. {Khalife} and Z.~M. {Kassas}, ``Receiver design for doppler positioning
  with leo satellites,'' in \emph{ICASSP 2019 - 2019 IEEE International
  Conference on Acoustics, Speech and Signal Processing (ICASSP)}, 2019, pp.
  5506--5510.

\bibitem{khalife2020blind}
J.~Khalife, M.~Neinavaie, and Z.~M. Kassas, ``Blind doppler estimation from leo
  satellite signals: A case study with real 5g signals,'' in \emph{Proceedings
  of the 33rd International Technical Meeting of the Satellite Division of The
  Institute of Navigation (ION GNSS+ 2020)}, 2020, pp. 3046--3054.

\bibitem{kassas2020navigation}
Z.~Z.~M. Kassas, ``Navigation from low-earth orbit,'' \emph{Position,
  Navigation, and Timing Technologies in the 21st Century: Integrated Satellite
  Navigation, Sensor Systems, and Civil Applications}, 2020.

\bibitem{kassas2020leo}
Z.~M. Kassas, \emph{Position, Navigation, and Timing Technologies in the 21st
  Century: Integrated Satellite Navigation, Sensor Systems, and Civil
  Applications}.\hskip 1em plus 0.5em minus 0.4em\relax Wiley-IEEE, 2020,
  vol.~1, ch. {N}avigation from {L}ow {E}arth {O}rbit: {P}art 2: {M}odels,
  {I}mplementation, and {p}erformance, pp. 1381--1412.

\bibitem{levanon1984theoretical}
N.~Levanon, ``Theoretical bounds on random errors in satellite {Doppler}
  navigation,'' \emph{IEEE Transactions on Aerospace and Electronic Systems},
  no.~6, pp. 810--816, 1984.

\bibitem{lawrence2017navigation}
D.~Lawrence, H.~Cobb, G.~Gutt, M.~OConnor, T.~Reid, T.~Walter, and D.~Whelan,
  ``Navigation from leo: Current capability and future promise,'' \emph{GPS
  World}, vol.~28, no.~7, pp. 42--48, 2017.

\bibitem{ardito2019performance}
C.~T. Ardito, J.~J. Morales, J.~Khalife, A.~Abdallah, Z.~M. Kassas
  \emph{et~al.}, ``Performance evaluation of navigation using {LEO} satellite
  signals with periodically transmitted satellite positions,'' in
  \emph{Proceedings of the 2019 International Technical Meeting of The
  Institute of Navigation}, 2019, pp. 306--318.

\bibitem{benzerrouk2019alternative}
H.~Benzerrouk, Q.~Nguyen, F.~Xiaoxing, A.~Amrhar, A.~V. Nebylov, and R.~Landry,
  ``Alternative pnt based on iridium next leo satellites doppler/ins integrated
  navigation system,'' in \emph{2019 26th Saint Petersburg International
  Conference on Integrated Navigation Systems (ICINS)}.\hskip 1em plus 0.5em
  minus 0.4em\relax IEEE, 2019, pp. 1--10.

\bibitem{Reid2016LeveragingCB}
T.~G.~R. Reid, A.~M. Neish, T.~Walter, and P.~K. Enge, ``Leveraging commercial
  broadband {LEO} constellations for navigation,'' in \emph{Proceedings of the
  29th International Technical Meeting of The Satellite Division of the
  Institute of Navigation (ION GNSS+ 2016)}, Portland, Oregon, Sept. 2016, pp.
  2300--2314.

\bibitem{Reid2018BroadbandLEO}
\BIBentryALTinterwordspacing
T.~G. Reid, A.~M. Neish, T.~Walter, and P.~K. Enge, ``Broadband {LEO}
  constellations for navigation,'' \emph{Navigation}, vol.~65, no.~2, pp.
  205--220, 2018. [Online]. Available:
  \url{https://onlinelibrary.wiley.com/doi/abs/10.1002/navi.234}
\BIBentrySTDinterwordspacing

\bibitem{reid2020leo}
T.~G.~R. Reid, T.~Walter, P.~K. Enge, D.~Lawrence, S.~Cobb, G.~Gutt,
  M.~O'Connor, and D.~Whelan, \emph{Position, Navigation, and Timing
  Technologies in the 21st Century: Integrated Satellite Navigation, Sensor
  Systems, and Civil Applications}.\hskip 1em plus 0.5em minus 0.4em\relax
  Wiley-IEEE, 2020, vol.~1, ch. {N}avigation from {L}ow {E}arth {O}rbit: {P}art
  1: {C}oncept, {C}apability, and {F}uture {P}romise., pp. 1359--1380.

\bibitem{SatNavXona2020}
T.~G.~R. Reid, B.~C. Chan, A.~Goel, K.~Gunning, B.~Manning, J.~Martin,
  A.~Neish, and A.~Perkins, ``Satellite navigation for the age of autonomy,''
  in \emph{Proceedings of the IEEE/ION PLANSx Meeting}.\hskip 1em plus 0.5em
  minus 0.4em\relax IEEE, 2020.

\bibitem{kassas2014receding}
Z.~M. Kassas and T.~E. Humphreys, ``Receding horizon trajectory optimization in
  opportunistic navigation environments,'' \emph{IEEE Transactions on Aerospace
  and Electronic Systems}, vol.~51, no.~2, pp. 866--877, April 2015.

\bibitem{kassas2014greedy}
Z.~M. Kassas, A.~Arapostathis, and T.~E. Humphreys, ``Greedy motion planning
  for simultaneous signal landscape mapping and receiver localization,''
  \emph{IEEE Journal of Selected Topics in Signal Processing}, vol.~9, no.~2,
  pp. 247 -- 258, March 2015.

\bibitem{kassas2020sops}
Z.~M. Kassas, \emph{Position, Navigation, and Timing Technologies in the 21st
  Century: Integrated Satellite Navigation, Sensor Systems, and Civil
  Applications}.\hskip 1em plus 0.5em minus 0.4em\relax Wiley-IEEE, 2020,
  vol.~1, ch. {N}avigation with {C}ellular {S}ignals of {O}pportunity, pp.
  1171--1224.

\bibitem{j_spilker96_dop}
J.~J. Spilker, \emph{Global Positioning System: Theory and Applications}.\hskip
  1em plus 0.5em minus 0.4em\relax Washington, D.C.: American Institute of
  Aeronautics and Astronautics, 1996, ch. 5: Satellite Constellation and
  Geometric Dilution of Precision, pp. 177--208.

\bibitem{psiaki2020navigation}
M.~L. Psiaki, ``Navigation using carrier doppler shift from a {LEO}
  constellation: {TRANSIT} on steroids,'' in \emph{Proceedings of the 33rd
  International Technical Meeting of the Satellite Division of The Institute of
  Navigation (ION GNSS+ 2020)}, 2020, pp. 3027--3045.

\bibitem{mclemore2020navigation}
B.~McLemore and M.~L. Psiaki, ``Navigation using doppler shift from leo
  constellations and ins data,'' in \emph{Proceedings of the 33rd International
  Technical Meeting of the Satellite Division of The Institute of Navigation
  (ION GNSS+ 2020)}, 2020, pp. 3071--3086.

\bibitem{starlink2021parameters}
SpaceX, ``Revised {SpaceX} gen2 non-geostationary satellite system, {Technical}
  {Attachment},''
  \url{https://licensing.fcc.gov/myibfs/download.do?attachment_key=12943362},
  Aug. 2021, {SAT-AMD-20210818-00105}.

\bibitem{kuiper2019parameters}
K.~Systems, ``Application for fixed satellite service by {Kuiper} {Systems}
  {LLC},''
  \url{https://licensing.fcc.gov/myibfs/download.do?attachment_key=1773885},
  July 2019, {SAT-LOA-20190704-00057}.

\bibitem{neinavaie2021exploiting}
M.~Neinavaie, J.~Khalife, and Z.~M. Kassas, ``Exploiting {Starlink} signals for
  navigation: First results,'' in \emph{Proceedings of the {ION} {GNSS}+
  Meeting}, St. Louis, Missouri, Sept. 2021, pp. 2766--2773.

\bibitem{starlink2020redditAma}
S.~engineering team, ``We are the {S}tarlink team, ask us anything!'' Nov.
  2020,
  \url{https://www.reddit.com/r/Starlink/comments/jybmgn/we_are_the_starlink_team_ask_us_anything/}.

\bibitem{starlinkgen2}
{Space Exploration Holdings}, ``{SpaceX} {Gen2} {NGSO} {Satellite} {System},
  {Attachment} {Waiver} {Requests},''
  \url{https://licensing.fcc.gov/myibfs/download.do?attachment_key=2378667},
  May 2020, {SAT-LOA-20200526-00055}.

\bibitem{starlink2019parameters}
SpaceX, ``{SpaceX} non-geostationary satellite system, {Technical}
  {Parameters},''
  \url{https://licensing.fcc.gov/myibfs/download.do?attachment_key=1877844},
  Aug. 2019, {SAT-MOD-20190830-00087}.

\bibitem{osoro2021techno}
O.~B. Osoro and E.~J. Oughton, ``A techno-economic framework for satellite
  networks applied to low earth orbit constellations: Assessing {Starlink,
  OneWeb and Kuiper},'' \emph{IEEE Access}, vol.~9, pp. 141\,611--141\,625,
  2021.

\bibitem{cid2015wideband}
E.~L. Cid, M.~G. Sanchez, and A.~V. Alejos, ``Wideband analysis of the
  satellite communication channel at {Ku}-and {X}-bands,'' \emph{IEEE
  Transactions on Vehicular Technology}, vol.~65, no.~4, pp. 2787--2790, 2015.

\bibitem{mostacciuolo2018modeling}
E.~Mostacciuolo, L.~Iannelli, S.~Sagnelli, F.~Vasca, R.~Luisi, and
  V.~Stanzione, ``Modeling and power management of a leo small satellite
  electrical power system,'' in \emph{2018 European Control Conference
  (ECC)}.\hskip 1em plus 0.5em minus 0.4em\relax IEEE, 2018, pp. 2738--2743.

\bibitem{perez2016power}
R.~P{\'e}rez-Torres, C.~Torres-Huitzil, and H.~Galeana-Zapi{\'e}n, ``Power
  management techniques in smartphone-based mobility sensing systems: A
  survey,'' \emph{Pervasive and Mobile Computing}, vol.~31, pp. 1--21, 2016.

\bibitem{faruqui2017arcturus}
A.~Faruqui, S.~Sergici, and C.~Warner, ``Arcturus 2.0: A meta-analysis of
  time-varying rates for electricity,'' \emph{The Electricity Journal},
  vol.~30, no.~10, pp. 64--72, 2017.

\bibitem{sen2012incentivizing}
S.~Sen, C.~Joe-Wong, S.~Ha, and M.~Chiang, ``Incentivizing time-shifting of
  data: A survey of time-dependent pricing for internet access,'' \emph{IEEE
  Communications Magazine}, vol.~50, no.~11, pp. 91--99, 2012.

\bibitem{murrian2021leo}
M.~J. Murrian, L.~Narula, P.~A. Iannucci, S.~Budzien, B.~W. O'Hanlon, S.~P.
  Powell, and T.~E. Humphreys, ``First results from three years of {GNSS}
  interference monitoring from low {Earth} orbit,'' \emph{Navigation, Journal
  of the Institute of Navigation}, vol.~68, no.~4, pp. 673--685, 2021.

\bibitem{teng2016closed}
Y.~Teng and J.~Wang, ``A closed-form formula to calculate geometric dilution of
  precision ({GDOP}) for multi-{GNSS} constellations,'' \emph{GPS Solutions},
  vol.~20, no.~3, pp. 331--339, 2016.

\bibitem{hobiger2013correction}
T.~Hobiger, D.~Piester, and P.~Baron, ``A correction model of dispersive
  troposphere delays for the aces microwave link,'' \emph{Radio Science},
  vol.~48, no.~2, pp. 131--142, 2013.

\bibitem{welsh1967upper}
D.~J. Welsh and M.~B. Powell, ``An upper bound for the chromatic number of a
  graph and its application to timetabling problems,'' \emph{The Computer
  Journal}, vol.~10, no.~1, pp. 85--86, 1967.

\bibitem{zufferey2008graph}
N.~Zufferey, P.~Amstutz, and P.~Giaccari, ``Graph colouring approaches for a
  satellite range scheduling problem,'' \emph{Journal of Scheduling}, vol.~11,
  no.~4, pp. 263--277, 2008.

\bibitem{montenbruck2005reduced}
O.~Montenbruck, T.~Van~Helleputte, R.~Kroes, and E.~Gill, ``Reduced dynamic
  orbit determination using {GPS} code and carrier measurements,''
  \emph{Aerospace Science and Technology}, vol.~9, no.~3, pp. 261--271, 2005.

\bibitem{montenbruck2021performance}
O.~Montenbruck, F.~Kunzi, and A.~Hauschild, ``Performance assessment of
  {GNSS}-based real-time navigation for the {Sentinel}-6 spacecraft,'' vol.~26,
  no.~12.\hskip 1em plus 0.5em minus 0.4em\relax Springer, Nov. 2021.

\bibitem{sun2017realtime}
\BIBentryALTinterwordspacing
X.~Sun, C.~Han, and P.~Chen, ``Precise real-time navigation of leo satellites
  using a single-frequency gps receiver and ultra-rapid ephemerides,''
  \emph{Aerospace Science and Technology}, vol.~67, pp. 228--236, 2017.
  [Online]. Available:
  \url{https://www.sciencedirect.com/science/article/pii/S1270963817306004}
\BIBentrySTDinterwordspacing

\bibitem{gettys2011bufferbloat}
\BIBentryALTinterwordspacing
J.~Gettys and K.~Nichols, ``Bufferbloat: Dark buffers in the internet: Networks
  without effective {AQM} may again be vulnerable to congestion collapse.''
  \emph{Queue}, vol.~9, no.~11, p. 40–54, Nov. 2011. [Online]. Available:
  \url{https://doi.org/10.1145/2063166.2071893}
\BIBentrySTDinterwordspacing

\bibitem{berrou1993turbo}
C.~Berrou, A.~Glavieux, and P.~Thitimajshima, ``Near shannon limit
  error-correcting coding and decoding: Turbo-codes. 1,'' in \emph{Proceedings
  of ICC '93 - IEEE International Conference on Communications}, vol.~2, 1993,
  pp. 1064--1070 vol.2.

\bibitem{buchsbaum1986pointing}
\BIBentryALTinterwordspacing
L.~M. Buchsbaum, ``Pointing losses in single-axis and fixed-mount earth-station
  antennas due to satellite movement,'' \emph{International Journal of
  Satellite Communications}, vol.~4, no.~2, pp. 89--96, 1986. [Online].
  Available:
  \url{https://onlinelibrary.wiley.com/doi/abs/10.1002/sat.4600040206}
\BIBentrySTDinterwordspacing

\bibitem{MuskTweetLatency}
\BIBentryALTinterwordspacing
E.~Musk, (@elonmusk), Twitter post: ``Around 20ms. {It's} designed to run
  real-time, competitive video games. {Version} 2, which is at lower altitude
  could be as low as 8ms latency.'' [Online]. Available:
  \url{https://twitter.com/elonmusk/status/1272363466288820224}
\BIBentrySTDinterwordspacing

\bibitem{broadbandNow2020unconnected}
\BIBentryALTinterwordspacing
J.~Busby and J.~Tanberk, ``{FCC} reports broadband unavailable to 21.3 million
  americans, {BroadbandNow} study indicates 42 million do not have access,''
  BroadbandNow Research, Tech. Rep., 2020. [Online]. Available:
  \url{https://broadbandnow.com/research/fcc-underestimates-unserved-by-50-percent}
\BIBentrySTDinterwordspacing

\bibitem{CIESIN_GPWv4_2018}
C.~for International Earth Science Information~Network, ``{Gridded Population
  of the World, Version 4 (GPWv4): National Identifier Grid, Revision 11.
  Palisades, NY: NASA Socioeconomic Data and Applications Center (SEDAC)},''
  Columbia University, Tech. Rep., 2018.

\bibitem{itu2019measuringFacts}
{International Telecommunications Union}, ``Measuring digital development:
  Facts and figures,'' International Telecommunications Union, Tech. Rep.,
  2019,
  \url{https://www.itu.int/en/ITU-D/Statistics/Documents/facts/FactsFigures2019.pdf}.

\bibitem{agrotis2017igsrts}
L.~Agrotis, E.~Sch{\"o}nemann, W.~Enderle, M.~Caissy, and A.~R{\"u}lke, ``The
  {IGS} {Real} {Time} {Service},'' in \emph{Proceedings of the 157th DVW
  Seminar}, Potsdam, Feb. 2017.

\bibitem{igs2020annual}
A.~Villiger and R.~Dach, ``International {GNSS} {Service} technical report 2020
  ({IGS} annual report),'' IGS Central Bureau and University of Bern Open
  Publishing, Tech. Rep., 2021.

\bibitem{sturze2020igsrts}
\BIBentryALTinterwordspacing
A.~St{\"u}rze, P.~Neumaier, and W.~S{\"o}hne, ``The {IGS} {Real} {Time}
  {Service}: Status and developments,'' in \emph{SIRGAS Symposium 2020}.
  [Online]. Available:
  \url{https://www.sirgas.org/fileadmin/docs/Boletines/Bol25/04_PeterNeumaier_SIRGAS.pdf}
\BIBentrySTDinterwordspacing

\end{thebibliography}

\end{document}